\documentclass[%
 aip,
 amsmath,amssymb,
 reprint,%
]{revtex4-1}

\usepackage{graphicx}
\usepackage{subcaption}
\usepackage{dcolumn}
\usepackage{bm}
\usepackage{comment}

\usepackage[utf8]{inputenc}
\usepackage[T1]{fontenc}
\usepackage{mathptmx}
\usepackage{etoolbox}

\makeatletter
\def\@email#1#2{%
 \endgroup
 \patchcmd{\titleblock@produce}
  {\frontmatter@RRAPformat}
  {\frontmatter@RRAPformat{\produce@RRAP{*#1\href{mailto:#2}{#2}}}\frontmatter@RRAPformat}
  {}{}
}%
\makeatother
\begin{document}

\preprint{AIP/123-QED}

\title[Variational Formulation of Reduced Kinetic Plasma]{Variational Formulation of a Hybrid Kinetic
and Gyrokinetic Model for Astrophysical and
Laboratory Plasmas}

\author{F. N. deOliveira-Lopes}
 \affiliation{MCentre for Mathematical Plasma Astrophysics, KU Leuven, Belgium}
\author{S. C. Thatikonda}
\affiliation{%
Max Planck Institute for Plasma Physics, Garching bei München,
Germany
}%
\author{D. Told }%
\affiliation{Max Planck Institute for Plasma Physics, Garching bei München,
Germany
}%
\author{A. Mustonen}
\affiliation{%
Max Planck Institute for Plasma Physics, Garching bei München,
Germany
}%
\author{K. Pommois }
\affiliation{%
Max Planck Institute for Plasma Physics, Garching bei München,
Germany
}%
\author{K. Hagiwara}
\affiliation{%
Max Planck Institute for Plasma Physics, Garching bei München,
Germany
}%
\author{F. Jenko}
\affiliation{%
Max Planck Institute for Plasma Physics, Garching bei München,
Germany
}%
\author{R. Grauer}
\affiliation{%
Ruhr-University Bochum, Bochum, Germany
}%

\date{\today}

\begin{abstract}
We present a Lagrangian model that encapsulates fully kinetic ions and gyrokinetic electrons for solar wind electromagnetic turbulence. Using a consistent method, where both electrons and protons are treated with the same mathematical formalism, we derive and implement a model in which high-frequency waves and kinetic electron effects are described in a computationally cost-efficient way. Higher-order Lie-transform perturbation methods applied to Hamiltonian formulation of guiding center motion are used to describe the dynamics of particles and fields. We use a variational approach to derive field equations for closure of the system.
\end{abstract}

\maketitle

\section{Introduction:\protect\\}\label{sec:level1}

Plasma turbulence is pervasive to astrophysics and space physics \citep{book_TurbulenceWind, book_SpacePhysics, book_PlasmaAstrophysics}. In weakly magnetized and weakly collisional plasmas, empirical data suggest that  kinetic effects play a pivotal role in understanding energy dissipation at ion and electron scales \citep{Alexandrova2009, TurbulenceDissipation, Chen2019}. Space plasma physics is idiosyncratically suitable for the understanding of turbulence because it offers a vast range of frequency scales and the possibility of \textit{in situ} measurements with various spacecraft missions \citep{Alexandrova2009, PSP_7, PSP_2, PSP_3}. Of these missions, the Parker Solar Probe is the most recent \citep{PSP_4, PSP_5}. Early data suggest that even at distances as close as 30 to 50 $R_\odot$, the transport of energy from large scales into thermal heat is already mediated through kinetic turbulent effects, as suggested by the power spectral density \citep{PSP_1, PSP_6, PSP_8}.

Energy dissipation at kinetic scales is a yet unresolved problem in space physics \cite{book_TurbulenceWind}. Currently, it is known that energy is injected into the system at the largest scale: it is then transported to the smallest scales, where it is transformed into thermal energy \citep{book_BasicAstro, TurbulenceDissipation}. In the process of dissipation, the statistically isotropic inertial range is elegantly described by magnetohydrodynamics \citep{MHD_1, MHD_2}. At the smallest scales, kinetic effects take place \citep{KAW_1}. Contrary to prior assumptions, the temperature of the solar wind at 1AU is higher than adiabatic expansion predicts \citep{Marsch, Kohl}; in this case, kinetic turbulent effects might be playing an important role.

Particle heating occurs in plasma physics due to irreversible thermodynamic processes and the breaking of time-reversal symmetry . The well-formulated Onsager reciprocity theorem \citep{Callen} together with the natural tendency of isolated systems to always evolve towards maximum entropy, allows us to better understand how energy is dissipated. In collisional plasma turbulence, the mechanism responsible for energy dissipation is viscosity \citep{Viscous}. The picture changes in the kinetic scales, where frictional conversion of the bulk energy does not completely explain how energy dissipation takes place. In this case, wave-particle interactions should be taken into account \citep{Carbone}. The wave-particle interaction forms structures in the velocity space distribution function \citep{Relaxation}, which in turn increase the efficacy of collisional energy dissipation \citep{Relaxation2}. Among the reversible energy transfer channels between waves and particle, the resonant effects take place when the particle gyrating around the magnetic field resonates with the wave's frequency, this can happen in the form of Landau damping \citep{pitaevskii2012physical} and cyclotron damping \citep{CyclotronDamping}. Non-resonant effects take place in the form of stochastic heating or magnetic pumping \citep{Stochastic, Pumping}. Whether relaxations in velocity space distribution function completely explain energy dissipation remains to be further investigated. Nevertheless, a more meticulous investigation of irreversible processes is needed in the framework of weakly collisional energy dissipation to properly understand the values of solar wind temperature measured \citep{Verscharen}. In moving forward, it is important to keep in mind previous work that aimed at modeling phenomena similar to the one we wish to investigate. The GK-electron and FK-ion (GKe/FKi) \citep{Lin_2005} was proposed to bridge the gap between fluid/kinetic hybrid models and fully kinetic models. Unlike the model proposed in this work, the parameter choices—such as the curvature of the magnetic field—in the literature surveyed are not well-suited for simulating turbulence in space and astrophysical plasmas. Furthermore, the derivation of the field equations follows a pull-back/push-forward  approach, whereas the present work uses a variational approach, which we expect will yield better conservation properties. 

To better investigate solar wind turbulence at high-frequency scales, we develop a theoretical framework focused on turbulent kinetic effects. We derive a hybrid system consisting of fully kinetic and gyrokinetic particle species. The gyrokinetic theory is an asymptotic limit of kinetic theory for magnetized plasmas. In this approach, a perturbation expansion is done to eliminate the gyro dependency up to a chosen ordering and to make sure that the magnetic momentum is conserved when the background magnetic field is stronger than the various  plasma perturbations. The present work is the first in a series that aims to build an electromagnetic nonlinear hybrid model that is both fully kinetic and gyrokinetic.


In the present work, we use a mathematical apparatus available in differential geometry and variational calculus. It is important to highlight here that the use of such tools not only allow us to to derive acceptable energy and momentum conservation properties. It also provides a route to ensure the preservation of the whole hierarchy of structures in the system. Beyond the conservation of energy and momentum, one is also capable of preserving the symplectic capacity, differential forms under Hamiltonian evolution, and conservation laws as derived from the Noether theorem. Moreover, it allows us to describe a system that organically encapsulates a geometric description of thermodynamics through contact geometry, the odd-dimensional counterpart of symplectic geometry \citep{Contact}. 

In 1958, F. E. Low published what would later be known as the first Lagrangian formulation of the Boltzmann-Vlasov equations \cite{Low}. In his work, he emphasized that, despite no new calculations being performed, the approach produced a new way to tackle practical problems. The contribution of the present work is a variationally derived hybrid kinetic model in which fully kinetic ions are coupled to gyrokinetic electrons through a common field formulation. The purpose is not to introduce a new wave or dissipation mechanism, but to provide a reduced model that retains selected electron-kinetic effects while extending the accessible frequency range relative to standard gyrokinetic treatments. We derive the particle and field equations consistently from the action, analyze the resulting electrostatic linear branches, and compare them with established kinetic and hybrid dispersion solvers.

This paper is organized as follows. In Section 2, we review the gyrokinetic coordinate transformation. In Section 3, we derive the Vlasov equations for fully kinetic ions and gyrokinetic electrons. Section 4 presents the derivation of the electromagnetic field equations using a variational formulation. Electrostatic linear results are analyzed in Section 5, while nonlinear electromagnetic results are discussed in Section 6. Finally, Section 7 summarizes the main findings and discusses their implications.

\section{Gyrokinetic coordinate derivation:\protect\\ Recapitulation}\label{sec:level2}

Rather than completely deriving the full gyrokinetic theory required for the present model, we only point out in this section the modifications implemented on the derivation performed in the literature, namely \cite{Littlejohn, littlejohn_1983, Tronko2018, Sugama}. A more detailed derivation of the present model can be found in \cite{Nathan}.

The standard gyrokinetic coordinate reduction consists of a series of mathematical transformations that aim to reduce the functional dependency of the system's Lagrangian on the gyration angle. This functional dependency is removed by eliminating the gyroangle \(\theta\) through perturbation theory.

The gyrokinetic transformation consists of a new ordering for the electric field, that is \((k_\perp \rho_{th}) \frac{e\phi_1}{T_i}= \varepsilon_\perp \varepsilon_\delta\). Furthermore, to prevent large fluctuations of the magnetic field, we also consider an ordering for the magnetic potential, that means \(\frac{\nabla A_1}{A_1} \sim \varepsilon_\delta^2\). For the full gyrokinetic derivation, \(\varepsilon_\perp \sim 1\), this condition can be modified in the case of a drift kinetic approximation. 

The first part of the transformation consists of the perturbation generated due to the symmetry break imposed by a background magnetic field. This symmetry transforms the fully kinetic Lagrangian 

\begin{equation*}
\Gamma=\left(m\textbf{v}+\frac{e}{c}\mathbf{A}(\textbf{x})\right)d\textbf{x} -Hdt,
\end{equation*}

where 

\begin{equation*}
H = \frac{m \textbf{v}^2}{2},
\end{equation*}

and $\mathbf{A}(\textbf{x})$, $\textbf{x}$, and $\textbf{v}$ represent the magnetic potential, position, and velocity coordinates, respectively. Since the strength of the symmetry break imposed by the background magnetic field is of a higher order, our model does not differ significantly from what is found in the literature up to this point. This means that, after the guiding center transformation, our Lagrangian retains the same format observed in \citep{Nathan}, i.e.

\begin{equation}
\begin{split}
\Gamma_{gc} = &\left[\frac{e}{c}A(X_{gc}) + mV_{\parallel}\hat{b}(X_{gc}) 
- \frac{mc}{e}\mu R \right] \boldsymbol{\cdot} dX_{gc} \\
&+ \frac{mc}{e}\mu d\Theta - H_{gc}dt
\end{split}
\label{full_lagrangian_1or2}
\end{equation}

where

\begin{equation}
H_{gc} =\frac{1}{2}mV_\parallel^{2}+\mu B.\label{some_H}
\end{equation}

Our symplectic tensor becomes then

\begin{equation}
\begin{split}
\omega_{gc} = d\Gamma_{gc} = &\frac{e}{c} \epsilon_{ijk} B^{k} dX_{gc}^{i} \wedge dX_{gc}^{j} \\
&+ m \hat{b}(X_{gc}) dV_\parallel \wedge dX_{gc}^{i} + \frac{mc}{e} d\mu \wedge d\theta
\end{split}
\label{gyrocenter_omega}
\end{equation}

and in matrix form, looks like 

\begin{equation*}
\omega_{gc} = \left(\begin{array}{cccccc}
0 & \frac{e}{c}B_{3} & -\frac{e}{c}B_{2} & - mb_{1} & 0 & 0\\
-\frac{e}{c}B_{3} & 0 & \frac{e}{c}B_{1} & - mb_{2} & 0 & 0\\
\frac{e}{c}B_{2} & -\frac{e}{c}B_{1} & 0 & - mb_{3} & 0 & 0\\
 mb_{1} & mb_{2} & mb_{3} & 0 & 0 & 0\\
0 & 0 & 0 & 0 & 0 & m\frac{c}{e}\\
0 & 0 & 0 & 0 & -m\frac{c}{e} & 0
\end{array}\right),
\end{equation*}

where the first three rows vs the first three columns represent \(dX_{gc}^{i}\wedge dX_{gc}^{j}\), the columns and rows with the terms \(mb_i\) represent \(dV_\parallel\wedge dX_{gc}^{i}\), and the terms with \(c/e\) represent the \(d\mu_{}\wedge d\theta_{}\) wedge product. To transform the Hamiltonian \(H_{gc} = \frac{1}{2}mV_\parallel^{2}+\mu B(\textbf{X}_{gc})\), we construct the vector field $X_{H_{gc}}$, determined by

\begin{equation}
i_{X_{H_{gc}}}\omega_{gc}+dH_{gc}=0,
\end{equation}

where $dH_{gc}$  is the exterior derivative of the Hamiltonian, a 1-form encoding how the energy changes in phase space, $X_{H_{gc}}$ is the Hamiltonian vector field associated with $H_{gc}$ , which generates the flow of the system in phase space, and $i_{X_{H_{gc}}}\omega_{gc}$  is the interior product (or contraction) of the symplectic form with the vector field $X_{H_{gc}}$ , yielding a 1-form.

The transformation of the symplectic form induces a change of the Poisson structure, and considering any two functions \(f,g\in(M_{gc},\mathbb{R})\) we have

\begin{equation}
\omega_{gc}(X_{f},X_{g})=\sum_{ab}\Pi_{gc}^{ab}(\Omega_{gc})\frac{\partial f}{\partial\Omega_{gc}^{a}}\frac{\partial g}{\partial\Omega_{gc}^{b}}.
\end{equation}

Since the relationship between the rank-2 Poisson tensor \(\Pi_{gc}=\omega_{gc}^{-1}\) also holds on our guiding-center manifold, we have our new Poisson bracket

\begin{multline}
\left\{ f,g\right\} _{gc}=\frac{e}{mc}\frac{1}{\epsilon_B}\left(\frac{\partial f}{\partial\theta}\frac{\partial g}{\partial\mu}-\frac{\partial g}{\partial\theta}\frac{\partial f}{\partial\mu}\right) \\
+\frac{B^{*}}{mB_{\parallel}^{*}}\cdot\left(\nabla^{*}f\frac{\partial g}{\partial V_{\parallel}}-\nabla^{*}g\frac{\partial f}{\partial V_{\parallel}}\right)-\epsilon_B\frac{c\hat{b}}{eB_{\parallel}^{*}}\cdot\left(\nabla^{*}f\times\nabla^{*}g\right) \label{gk_Poisson_bracket},
\end{multline}

where the modified gradient is given by \(\nabla^{*}=\nabla-\boldsymbol{R}^{*}\partial_{\theta_{}}\), and \(\boldsymbol{B}^{*}=\boldsymbol{B}+\frac{m}{e}cv_{\parallel}\nabla\times\hat{\boldsymbol{b}}-\frac{mc^{2}}{e^{2}}\mu\nabla\times \boldsymbol{R}^{*}\). To compute the final guiding center coordinate transformation, we need to find the generator \(S_{gc}\), which can be constructed in such a way that eliminates the fluctuations on \(\tilde{H} = H - \left \langle H \right \rangle\). The Hamiltonian transformation is given by the Lie transformation

\begin{equation*}
H_{gc}(Z_{gc})=e^{-\mathcal{L}_{Sgc}}H(Z),
\end{equation*}

where 

\begin{equation}
\mathcal{L}_{S_{gc}} = \{ S_{gc}, \boldsymbol{\cdot} \}_{gc}.\label{Lie}
\end{equation}

Equation \ref{Lie}, together with the Hamiltonian, should satisfy the cohomological equation

\begin{equation}\label{cohomological_equation}
\frac{eB}{mc}\frac{\partial S_{gc}}{\partial\theta}=(mV_{\parallel}\tilde{\mathcal{V}}_{\parallel}+mV_{\perp}\tilde{\mathcal{V}}_{\perp}),
\end{equation}

where  $\tilde{\mathcal{V}}_{\parallel}$ and $\tilde{\mathcal{V}}_{\perp}$ are the fluctuating part of $V_{\parallel}$ and $V_{\perp}$. That gives us 

\begin{align*}
S_{gc} &= \frac{m^{3}c^{2}}{e^{2}B^{2}} \bigg[
\frac{V_{\parallel}V_{\perp}^{2}}{8} \Big(
(\hat{a} \boldsymbol{\cdot} \nabla) \hat{b} \boldsymbol{\cdot} \hat{a}
- (\hat{c} \boldsymbol{\cdot} \nabla) \hat{b} \boldsymbol{\cdot} \hat{c}
\Big) \notag \\
&\quad + V_{\parallel}^{2}V_{\perp} (\nabla \times \hat{b}) \boldsymbol{\cdot} \hat{a} 
+ \frac{V_{\perp}^{3}}{3B} \hat{c} \boldsymbol{\cdot} \nabla B
\bigg].
\end{align*}

It is important to emphasize that, before solving equation \ref{cohomological_equation}, one must first apply the fluctuating operation to the given coordinate, i.e. \(\tilde{Z}_{gc} =  Z_{gc} - \left \langle Z_{gc} \right \rangle\), within the Hamiltonian. The coordinate transformation now solely affects the Hamiltonian's structure, implying that we do not anticipate structural alterations in the Lagrangian. Combined with the latest Lie transformation, the new coordinate system takes the form

\begin{equation*}
z_{gc}=Z_{gc}+\sum_{n=1}\frac{\epsilon_{B}^{n}}{n!}\mathcal{Z}_{n,gc}(Z_{gc})+\left\{ S_{gc},Z_{gc}\right\} _{gc},
\end{equation*}

where our Poisson bracket is constructed from the newly transformed symplectic component of the Lagrangian, and

\begin{equation*}
z_{gc}=\left[\begin{array}{c}
\textbf{x}_{gc}\\
\mathbf{v}_{s,\parallel}\\
\mathbf{v}_{s,\perp}\\
\theta
\end{array}\right], \quad
Z_{gc}=\left[\begin{array}{c}
X_{gc}\\
V_{\parallel}\\
V_{\perp}\\
\Theta
\end{array}\right], \quad
\mathcal{Z}_{gc}=\left[\begin{array}{c}
\chi\\
\mathcal{V}_{\parallel}\\
\mathcal{V}_{\perp}\\
\Omega
\end{array}\right],
\end{equation*}

\begin{equation*}
\{S_{gc},Z_{gc}\}=\left[\begin{array}{c}
(\hat{b}\boldsymbol{\cdot}\chi)\hat{b}-\frac{\hat{b}}{m}\frac{\partial S_{gc}}{\partial V_{\parallel}}\\
0\\
\frac{eB}{m^{2}V_{\perp}c}\frac{\partial S_{gc}}{\partial\Theta}\\
\frac{eB}{m^{2}V_{\perp}c}\frac{\partial S_{gc}}{\partial V_{\perp}}
\end{array}\right],
\end{equation*}

up to the ordering of our derivation. The second part of the coordinate transformation consists of the addition of the symmetry break due to the presence of electromagnetic perturbation. This transformation gives us our final gyrocenter coordinate transformation.

After a velocity shift and a Lie transformation, a new generating function is required to move towards the new coordinate system.

A few modifications to the standard derivations are necessary. More details of this modification can be found elsewhere, see \citep{Nathan}. To be concise, here we will content ourselves with setting out and understanding the reason behind the modifications.

In constructing the new Hamiltonian, three Poisson brackets are required to find the equation for the first order generating function \(S_1\) of the gyrokinetic Lie transformation, these are

\begin{equation}
\{S_{1},H_{1}\}_{f}=\frac{e}{mc}\left(\frac{\partial S_{1}}{\partial\theta}\frac{\partial H_{1}}{\partial\mu}-\frac{\partial H_{1}}{\partial\theta}\frac{\partial S_{1}}{\partial\mu}\right),
\end{equation}

\begin{equation}
\{S_{1},H_{1}\}_{m}=\frac{B^{*}}{mB_{\parallel}}\left(\nabla^{*}S_{1}\frac{\partial H_{1}}{\partial v_{\parallel}}-\nabla^{*}H_{1}\frac{\partial S_{1}}{\partial v_{\parallel}}\right)=\mathcal{O}(\varepsilon_{\delta}^{3}),
\end{equation}

and

\begin{equation}
\{S_{1},H_{1}\}_{s}=\frac{c\hat{b}}{eB_{\parallel}^{*}}\left(\nabla^{*}S_{1}\times\nabla^{*}H_{1}\right)
\end{equation}

where \(H_1\) stands for the perturbed Hamiltonian, and the indexes f, m, and s stands for fast, medium and slow motion, based on the ordering of each term. In our model, due to the difference in strength of the fields acting on the gyrokinetic species vis-à-vis the fully kinetic species, we can consider the last term equals to zero. This change propagates in the derivation and the final Lagrangian becomes

\begin{align*}
\Gamma_{gy} = &\left(\frac{e}{\varepsilon_\delta c} \textbf{A}_{1} 
+ m\textbf{v}_{gy, \parallel }\hat{\textbf{b}}(X_{gy})\right)
\boldsymbol{\cdot} \dot{\textbf{X}}_{gy} 
+ \varepsilon_\delta \frac{mc}{e} \mu_{gy} \dot{\theta}_{gy} \\
&- \frac{1}{2} m v_{gy, \parallel}^{2} 
- \mu_{gy} B(\textbf{X}_{gy}) 
- \varepsilon_{\delta} e \langle\psi_{1}\rangle \\
&- \varepsilon_{\delta}^{2} e^{2} \left(
\frac{1}{2mc^{2}} \langle|\textbf{A}_{1}|^{2}\rangle 
- \frac{1}{2B(\textbf{X}_{gy})} \partial_{\mu_{gy}} \langle\psi_{1}^{2}\rangle
\right)
\end{align*}

where our new coordinate system is described by

\begin{equation}
Z_{gy}=z_{gc}+\varepsilon_{\delta}\left\{ S_{gy},z_{gc}\right\} _{gc},
\end{equation}

where, up to the ordering of our derivation, we have that

\begin{equation*}
Z_{gy}=\left[\begin{array}{c}
X_{gy}\\
v_{gy,\parallel}\\
\mu_{gy}\\
\theta_{gy}
\end{array}\right], \quad
z_{gc}=\left[\begin{array}{c}
X_{gc}\\
v_{s,\parallel}\\
\mu\\
\theta
\end{array}\right],
\end{equation*}

\begin{equation*}
\{S_{gy},Z_{gc}\}=\left[\begin{array}{c}
-\left(\frac{c\hat{b}}{eB_{\parallel}^{*}}\times\nabla S_{1}+\frac{B^{*}}{mB_{\parallel}^{*}}\frac{\partial}{\partial v_{s,\parallel}}\int^{\theta}\tilde{H}_{1}d\theta\right)\\
0\\
\frac{e}{B}\tilde{H}_{1}\\
-\frac{1}{B}\frac{\partial}{\partial\mu}\int^{\theta}\tilde{H}_{1}d\theta
\end{array}\right].
\end{equation*}

Here we summarize the modifications outlined in \citep{Nathan} while some older literature such as \citep{Littlejohn} focuses solely on a guiding center approach, our method incorporates more recent strategies by accounting for electromagnetic perturbation fields through the gyrokinetic transformation. Similarly to \citep{Tronko2018}, we employ an ordering related to the curvature of the background magnetic field. This is in contrast to Littlejohn's approach \citep{Littlejohn}, which assumes an arbitrary \(\epsilon\) and conducts the asymptotic derivation as \(\epsilon \rightarrow 0\). Additionally, our model considers a slab background magnetic field, further distinguishing it from \citep{Tronko2018}.

\section{Equations of Motion:\protect\\Ion and Electron Vlasov equation}\label{sec:level3}

\subsection{Ion Vlasov Equation}

For the sake of coherence, we derive the equations for the fully kinetic species within the same framework as the gyrokinetic reduction. Starting with a tautological one-form \cite{Was} $\theta = \Gamma$, where $\Gamma=\mathcal{L}dt$ , the Poincaré two-form, also known as the canonical symplectic form, can then be expressed by the symplectization of $\theta$

\begin{equation}
    \omega=dp_{i}\wedge dq^{i},
\end{equation}

where

\begin{equation}
    p_i = p_i(\Theta^i,t),
\end{equation}

\begin{equation}
    q^i = q^i(\Theta^i,t),
\end{equation}

with $\Theta^a = (x,v)$, the non canonical phase space, with the transformations $x=q$. From the canonical momentum we have $v=p-\frac{e}{c}A$. We can write our non canonical Lagrangian one-form as 

\begin{equation}\label{first_lagrangian}
    \mathcal{L} = \Lambda_a \frac{\partial \Theta^a }{\partial t} - \bar{H},
\end{equation}\label{fully_L}

where $\Lambda_a = p_i \frac{\partial q^i}{\partial \Theta^a}$, and $\bar{H} = p_i \frac{\partial q^i }{\partial t} - H(\Theta) = \frac{1}{2m} |p - \frac{e}{c}A(q,t)|^2 - e\phi(q,t)$. 
Explicitly, $p=mv+\frac{e}{c}A(x,t)$, and with the appropriate substitutions, our tautological one-form $\Gamma = \mathcal{L}dt = (mv+\frac{e}{c}A(x,t) \cdot dx  - {H(x,v)}dt)$, becomes our symplectic form

\begin{equation}\label{symplectic_matrix}
\omega = d\Gamma = \epsilon_{ijk}B_{k}dx^{i}\wedge dx^{j}-m\delta_{ij}dx^{i}\wedge dv^{j}+m\delta_{ij}dv^{i}\wedge dx^{j}. 
\end{equation} \\

The characteristics of our Vlasov equation are computed with a Poisson bracket, which requires a Poisson tensor defined as 

\begin{equation*}
\begin{split}\Pi^{ij}=\omega_{ij}^{-1}=\begin{pmatrix}0 & \frac{1}{m}\delta^{ij}\\
-\frac{1}{m}\delta^{ij} & \frac{1}{m^{2}}\epsilon^{ijk}B_{k}
\end{pmatrix}\end{split},
\end{equation*}

and our Poisson bracket becomes 

\begin{equation}
\begin{split}
\{f,g\} &= \sum_{ij} \frac{\partial f}{\partial \Theta^{i}} \Pi^{ij} \frac{\partial g}{\partial \Theta^{j}} \\
&= \frac{1}{m} \left(\nabla_x f \cdot \nabla_v g
- \nabla_v f \cdot \nabla_x g \right) +\frac{1}{m^2}B \left(\nabla_v f \times \nabla_v g
 \right).
\end{split}
\label{poisson_bracket}
\end{equation}

Here $\nabla_x$ and $\nabla_v$, denote the gradients with respect to position and velocity, respectively. To construct the Vlasov equation, we should evaluate  Liouville’s theorem  in the context of symplectic geometry\cite{Lanczos}. Considering that the phase-space distribution function $F(p,q)$ is constant along the trajectories of the system, we can describe the time evolution of a volume element of our phase space according to the following relation

\begin{equation*}
    \frac{dF}{dt}=\frac{\partial F}{\partial t}+\sum_{i=1}^{n}\left(\frac{\partial F}{\partial q^{i}}\dot{q}^{i}+\frac{\partial F}{\partial p_{i}}\dot{p}_{i}\right)=0.
\end{equation*}

Because the phase space distribution function is constant along trajectories in the phase space, the theorem says that the Liouville measure is invariant under Hamiltonian flows. The relationship becomes 

\begin{equation*}
    \frac{dF}{dt}= \{F ,H \},\label{liouville}
\end{equation*}

and the Vlasov equation becomes  

\begin{equation*}
\begin{split}\frac{\partial F^{}}{\partial t}+\left\{ x^{},H^{}\right\} \cdot \nabla F^{}+\left\{ v^{},H^{}\right\} \partial_{v^{}}F^{}=0.\end{split}
\end{equation*}

Substituting the values derived from equation \ref{poisson_bracket} using $(x,v)$ as our coordinate system, we have the distribution function equation for the fully kinetic species

\begin{equation}
\frac{\partial F_{}^{}}{\partial t}+v \cdot \nabla_x F+ \left(\frac{\partial\phi}{\partial x^{i}}+\epsilon^{ijk}v_{j}B_{k}\right)\partial_{v^{i}}F_{}^{}=0.\label{final_fully}
\end{equation}

Finally, it is worth highlighting that the present approach offers a promising foundation for extending the framework to entropy-related physics, particularly in the context of metriplectic dynamics \cite{metriplectic}.

\subsection{Electron Vlasov Equation}

The derivation of the electron Vlasov equation starts with the Lagrangian derived in section \ref{sec:level2}, that is 

\begin{align*}
\Gamma_{gy} = &\left(\frac{e}{\varepsilon_\delta c} \textbf{A}_{1} 
+ m\textbf{v}_{gy, \parallel }\hat{\textbf{b}}(X_{gy})\right)
\boldsymbol{\cdot} \dot{\textbf{X}}_{gy} 
+ \varepsilon_\delta \frac{mc}{e} \mu_{gy} \dot{\theta}_{gy} \\
&- \frac{1}{2} m v_{gy, \parallel}^{2} 
- \mu_{gy} B(\textbf{X}_{gy}) 
- \varepsilon_{\delta} e \langle\psi_{1}\rangle \\
&- \varepsilon_{\delta}^{2} e^{2} \left(
\frac{1}{2mc^{2}} \langle|\textbf{A}_{1}|^{2}\rangle 
- \frac{1}{2B(\textbf{X}_{gy})} \partial_{\mu_{gy}} \langle\psi_{1}^{2}\rangle
\right).
\end{align*}

From this Lagrangian, we can compute the characteristics and implement them on our modified Poisson bracket (Eq. \ref{gk_Poisson_bracket}.) The comprehensive gyrokinetic coordinate reduction used in the present model is elaborated in detail in \cite{Nathan}. 

Following the same procedure applied to the ion Vlasov equations, we derive the gyrokinetic equations of motion and obtain the final set of equations

\begin{equation}
\dot{\textbf{X}}_{gy}=\left\{ \textbf{X}_{gy},H_{gy}\right\} =\frac{B^{*}}{mB_{\parallel}^{*}}\frac{\partial H_{gy}}{\partial v_{gy, \parallel}}+\varepsilon_\delta \frac{c\hat{b}}{eB_{\parallel}^{*}} \times \nabla^{*}H_{gy},
\end{equation}
\begin{multline}
\dot{v}_{gy, \parallel} =\frac{\textbf{B}^{*}}{m{B}_{\parallel}^{*}}\boldsymbol{\cdot}\left(\nabla^*v_{gy, \parallel}\frac{\partial H_{gy}}{\partial v_{gy, \parallel}}-\nabla^{*}H_{gy}\right) \\
-\frac{c\hat{{b}}}{e{B}_{\parallel}^{*}}\varepsilon_\delta\left(\nabla^{*}v_{gy, \parallel}\times\nabla^{*}H_{gy}\right) = -\frac{\textbf{B}^{*}}{m{B}_{\parallel}^{*}}\boldsymbol{\cdot}\left(\nabla^{*}H_{gy}\right),
\end{multline}

\begin{equation}
\begin{split}
\dot{\mu}_{gy} &= \left\{ \mu_{gy}, H_{gy} \right\} 
= -\frac{e}{mc} \frac{1}{\varepsilon_\delta} \partial_{\theta_{gy}} H_{gy} \\
&\quad + \frac{\textbf{B}^{*}}{m{B}_{\parallel}^{*}} \nabla \mu_{gy} - \frac{c\hat{\textbf{b}}}{e{B}_{\parallel}^{*}} 
\varepsilon_\delta \left(\nabla \mu_{gy} \times \nabla^{*} H_{gy}\right)
\end{split}
\label{eq:20}
\end{equation}

which give us in turn

\begin{equation}
\dot{\mu}_{gy}=-\frac{e}{mc}\frac{1}{\varepsilon_\delta}\partial_{\theta_{gy}}H_{gy}=0\label{eq:21},
\end{equation}
and, lastly,

\begin{equation}
\dot{\theta}_{gy}=\frac{e}{mc}\frac{1}{\varepsilon_\delta}\partial_{\mu_{gy}}H_{gy}+\nabla^{*}\theta_{gy}
\boldsymbol{\cdot}\textbf{X}_{gy},
\end{equation}

where, the second term represents the implicit time dependence of the gyroangle through the gyrocenter trajectory. It is crucial to note that for the present model, only $\dot{\textbf{X}}_{gy}$ and $\dot{v}_{gy, \parallel} $ are needed to compute the dynamics of the system, up to the chosen ordering. Furthermore, since our system is derived to be theta-independent, there was no need to compute the equation of motion for this variable, which are shown here to demonstrate that, as expected, both those variables do not influence the dynamics of our system.

Our gyrokinetic Vlasov equation becomes

\begin{equation*}
   \frac{\partial F}{\partial t}+\frac{\textbf{B}^{*}}{m{B}_{\parallel}^{*}}\boldsymbol{\cdot}\left(mv_{gy, \parallel}+\frac{e}{c}\left\langle {A}_{1\parallel}\right\rangle \right)\nabla_{gy}F \\
   \end{equation*}
   \begin{equation*}
   +\frac{c\hat{\textbf{b}}}{e{B}^{*}_{\parallel}}\times\left(\mu_{gy}\nabla_{gy}B(\textbf{X}_{gy})-\varepsilon_{\delta}e\nabla\left\langle \phi_{1}\right\rangle +\varepsilon_{\delta}\frac{e}{c}v_{gy, \parallel}\left\langle \nabla {A}_{1\parallel}\right\rangle \right) \boldsymbol{\cdot} \nabla_{gy}F \\
   \end{equation*}
   \begin{equation*}
   -\frac{\textbf{B}^{*}}{m{B}_{\parallel}^{*}} \boldsymbol{\cdot} \left(\mu_{gy}\nabla_{gy}B(\textbf{X}_{gy})-\varepsilon_{\delta}e\nabla\left\langle \phi_{1}\right\rangle +\varepsilon_{\delta}\frac{e}{c}v_{gy, \parallel}\left\langle \nabla {A}_{1\parallel}\right\rangle \right)\partial_{v_{gy, \parallel}}F \\
\end{equation*}
   \begin{equation}
          =0,
   \end{equation}

where the second term on the first line is associated with the canonical momentum, while the terms on the second line are the grad B drift, $E \times B$ and magnetic flutter. The last term is associated with the total parallel acceleration.

\section{Fields:\protect\\Derivation of the full electromagnetic system}\label{sec:level4}

\subsection{Poisson}

To derive the field equations governing our system, we begin by considering that the system exists on a heterogeneous manifold \cite{heterogeneous_manifolds}. Specifically, the ions are described by their full kinetic dynamics in phase space $\Omega = (\mathbf{x}, \mathbf{v})$, while the electrons are treated using gyrokinetic theory, which involves a reduction in the coordinate space by averaging over the fast gyro-motion, leading to the phase space $\Omega_{gy} = (\mathbf{X}_{gy}, v_{\parallel gy}, \mu_{gy}, \theta)$.

In our derivation, the electromagnetic field serves as the bridge between these two descriptions, the fully kinetic ions and the gyrokinetic electrons, since it interacts with both species. We assume that the field Lagrangian is defined within the same manifold as the fully kinetic particles. This assumption allows us to consistently couple the electromagnetic fields with the fully kinetic ion dynamics while incorporating the effects of the fields on the gyrokinetic electrons through appropriate transformations.

To derive the field equations, we employ the variational principle. Specifically, we consider the variation of the action $\mathcal{S}$ with respect to a field $\chi(\Omega)$, yielding the condition for extremizing the action:

\begin{equation}
    \frac{\delta\mathcal{S}[\chi(\Omega)]}{\delta\chi(\Omega)}\circ\hat{\chi}(\Omega)=0.
\end{equation}

In this expression, $\mathcal{S}$ denotes the action of the system, $\chi(\Omega)$ is the field over which we perform the variation, and $\hat{\chi}(\Omega)$ is an arbitrary test field used to probe the variations. For a detailed derivation of the functional derivative in the context of heterogeneous manifolds, we refer the reader to \cite{Was, heterogeneous_manifolds}.

Since our system comprises both fully kinetic ions and gyrokinetic electrons, their respective phase spaces are different. The electromagnetic field couples to both species but is naturally defined in the physical coordinate space $\mathbf{x}$. Therefore, to perform the variation consistently, we need to transform quantities defined in the gyrokinetic coordinates back to the physical coordinates. This transformation allows us to carry out the variation within a common coordinate system. For instance, we have:

\begin{equation*}
\frac{\delta\phi_{1}(\mathbf{X}_{gy}+\rho)}{\delta\phi_{1}(\mathbf{x})}\circ\hat{\chi}(\mathbf{x})=\hat{\chi}(\mathbf{X}_{gy}+\rho).
\end{equation*}

Here, $\phi_{1}(\mathbf{X}_{gy}+\rho)$ represents the scalar potential evaluated at the position $\mathbf{X}_{gy}+\rho$, where $\rho$ is the gyroradius vector, and we express its variation in terms of the physical coordinate $\mathbf{x}$.

We proceed by considering the Lagrangian $\mathcal{L}$ introduced in previous sections, which together comprises contributions from the gyrokinetic electrons and the fully kinetic ions. Furthermore, we add here the electromagnetic field Lagrangian, which give us then

\begin{equation}
\mathcal{L} =  \mathcal{L}_{gy}^{p} + \mathcal{L}_{fk,i}^{p} + \mathcal{L}^{f},
\end{equation}

where $\mathcal{L}_{gy}^{p}$ is the particle Lagrangian for the gyrokinetic electrons, $\mathcal{L}_{fk,i}^{p}$ is the particle Lagrangian for the fully kinetic ions, and $\mathcal{L}^{f}$ is the field Lagrangian. The total action $\mathcal{S}$ is then obtained by integrating the Lagrangian over time. For multiple ion species, the action can be expressed as:

\begin{equation}
\mathcal{S} = \mathcal{S}_{gy}^{p} + \sum_i \mathcal{S}_{fk,i}^{p} + \mathcal{S}^{f},
\end{equation}
where the individual components of the action are given by:

For the gyrokinetic electrons:
\begin{multline}
\mathcal{S}_{gy}^{p} = \int \int F_e(\mathbf{X}_{gy}, v_{gy, \parallel}, \mu_{gy}, t) \left\{ \left( \frac{-e}{\varepsilon_\delta c} \mathbf{A}_{0} + m v_{gy, \parallel} \hat{b}(\mathbf{X}_{gy}) \right) \cdot \dot{\mathbf{X}}_{gy} \right. \\ 
\left. - \varepsilon_\delta \frac{m c}{e} \mu_{gy} \dot{\theta}_{gy} - \frac{1}{2} m v_{gy, \parallel}^{2} - \mu_{gy} B(\mathbf{X}_{gy}) + \varepsilon_\delta e \langle \psi_{1} \rangle \right. \\ 
\left. - \varepsilon_\delta^{2} e^{2} \left( \frac{1}{2 m c^{2}} \langle | \mathbf{A}_{1} |^{2} \rangle - \frac{1}{2 B(\mathbf{X}_{gy})} \partial_{\mu_{gy}} \langle \tilde{\psi}_{1}^{2} \rangle \right) \right\} \, dtd\Omega_{gy},
\end{multline}
where $\psi_1 \equiv \phi_1 - (v_{\parallel gy}/c)A_{1\parallel}$ is the generalized potential, $\langle \cdot \rangle$ denotes the gyroaverage at fixed $(\mathbf{X}_{gy}, \mu_{gy})$, and  $ \tilde\psi_1 \equiv \psi_1 - \langle \psi_1\rangle$ is the gyrophase-dependent part.

For the fully kinetic ions:
\begin{multline}
\mathcal{S}_{fk, i}^{p} = \int \int F_i(\mathbf{x}, \mathbf{v}) \left\{ \left( m_{i} \mathbf{v} + \frac{q_i}{c} [ \mathbf{A}_{0}(\mathbf{x}) + \varepsilon_{\delta} \mathbf{A}_{1}(\mathbf{x}) ] \right) \cdot \dot{\mathbf{x}} \right. \\
\left. - \frac{1}{2} m_{i} |\mathbf{v}|^{2} - q_i \phi(\mathbf{x}) \right\} \, dtd\Omega,
\end{multline}
and for the electromagnetic fields

\begin{multline}
\mathcal{S}^{f} = \int \left\{ \frac{1}{8\pi} \int d^{3}x \left( \varepsilon_{\delta}^{2} | \nabla \phi_{1}(\mathbf{x}) |^{2} \right. \right. \\
\left. \left. - | \nabla \times [ \mathbf{A}_{0}(\mathbf{x}) + \varepsilon_{\delta} \mathbf{A}_{1}(\mathbf{x}) ] |^{2} \right) \right\} dt.
\end{multline}

In these expressions, $F_e$ and $F_i$ are the distribution functions for the electrons and ions, respectively; $e$ and $q_i$ are the electron and ion charges; $m$ and $m_i$ are the electron and ion masses; $\mathbf{A}_{0}$ and $\mathbf{A}_{1}$ are the zeroth and first-order magnetic potentials; $\phi_{1}$ is the first-order scalar potential; $\varepsilon_\delta$ is a small parameter representing the order of the perturbation; $\hat{b}$ is the unit vector along the magnetic field; $\langle \cdot \rangle$ denotes the gyroaverage; and $\tilde{\psi}_{1}$ is the fluctuating part of the potential. The variables $\mathbf{X}_{gy}$, $v_{gy, \parallel}$, $\mu_{gy}$, and $\theta_{gy}$ are the gyrokinetic coordinates. 

By varying the action $\mathcal{S}$ with respect to the scalar potential $\phi_{1}$, we derive the Poisson equation for our system. This involves integrating over the appropriate phase spaces and accounting for the contributions from both the gyrokinetic electrons and the fully kinetic ions.

Here $\Psi_1(\mathbf X) \equiv \langle \phi_1(\mathbf X + \boldsymbol\rho)\rangle$ denotes the gyroaveraged electrostatic potential entering the gyrokinetic electron dynamics. After performing the variation and simplifying the resulting expressions, we obtain the Poisson equation
\begin{equation*}
\frac{1}{4\pi}\nabla ^2 \phi_{1}(\mathbf{x}) +  \frac{\rho_{th}^{2}}{\lambda_{D}^{2}} \nabla_{\perp}^2 \Psi_{1}(\mathbf{x}) = \sum_i q_i n_i(\mathbf{x}) - e n_e(\mathbf{x}),
\end{equation*}
where $\rho_{th}$ is the electron thermal gyroradius, $\lambda_{D}$ is the electron Debye length, and $\Psi_{1}(\mathbf{x})$ represents the first-order potential. The electron density $n_e$ is the gyrocenter density, while $n_i$ denotes the particle density of the fully kinetic ions. The term involving $\nabla_{\perp}^2 \Psi_{1}$ results from the long-wavelength expansion of the gyrokinetic polarization density, it is the gyrokinetic polarization due to finite-Larmor-radius effects.

This equation can be derived by recognizing that the integration over the gyrokinetic variables yields the electron density $n_e$ via $\int dW F_e(\mathbf{x})= n_e$, and by utilizing the relationship
\[
\frac{m c^{2}}{2 B^{2}(\mathbf{X}_{gy})} n_e=\frac{1}{8\pi} \frac{\omega_{pe}^2}{\Omega_{ce}^2} = \frac{1}{8\pi}\frac{\rho_{th}^{2}}{\lambda_{D}^{2}},
\]
where $\omega_{pe}$ is the electron plasma frequency and $\Omega_{ce}$ is the electron cyclotron frequency.

To elucidate the relationship between the Poisson equation and the parallel component of Ampère's equation, we express the electromagnetic potentials explicitly. Considering the contributions from both the scalar potential $\phi_{1}$ and the vector potential $\mathbf{A}_{1}$, we rewrite the Poisson equation as

\begin{equation}
\begin{split}
 \frac{1}{4\pi} \nabla_\perp^2 \phi_{1}(\mathbf{x}) 
&+  \frac{\rho_{th}^{2}}{\lambda_{D}^{2}} 
\left(  \nabla_{\perp}^2 \phi_{1}(\mathbf{x}) 
- \frac{u_{e \parallel}}{c}\nabla_\perp^2 A_{1\parallel}(\mathbf{x}) \right) \\
&= \sum_i q_i n_i(\mathbf{x}) - e n_e(\mathbf{x}),
\end{split}
\label{final_ampere_eq}
\end{equation}

This expression highlights the coupling between the scalar and vector potentials in the presence of gyrokinetic effects.

Rearranging terms and factoring out $\nabla_{\perp}^{2} \phi_{1}$, we can write the equation as:
\begin{equation}
\begin{split}
\left( \frac{1}{4\pi} + \frac{\rho_{th}^{2}}{\lambda_{D}^{2}} \right) 
\nabla_{\perp}^{2}\phi_{1}(\mathbf{x}) 
&- \frac{u_{e \parallel}}{c} \frac{\rho_{th}^{2}}{\lambda_{D}^{2}} 
\nabla_{\perp}^2 A_{1 \parallel}(\mathbf{x}) \\
&= \sum_i q_i n_i(\mathbf{x}) - e n_e(\mathbf{x}),
\end{split}
\label{modified_ampere_eq}
\end{equation}

where $A_{1 \parallel}$ is the parallel component of the first-order vector potential, and $u_{e \parallel}=\int v_{\parallel gy}  F_e\,dW$ is the parallel electron flow velocity. Alternatively, by combining the coefficients, we can express it as
\begin{equation}
\begin{split}
\frac{1}{4\pi} \nabla_{\perp}^{2}\phi_{1} 
\left( 4\pi\frac{\rho_{th}^{2}}{\lambda_{D}^{2}} + 1   \right) 
&-\frac{u_{e \parallel}}{c} \frac{\rho_{th}^{2}}{\lambda_{D}^{2}} 
\nabla_{\perp}^2 A_{1 \parallel}(\mathbf{x}) \\
&= \sum_i q_i n_i(\mathbf{x}) - e n_e(\mathbf{x}),
\end{split}
\label{modified_ampere_eq}
\end{equation}

Assuming perpendicular gradients of $u_{e\parallel}$ are negligible at the ordering retained, $\nabla_\perp^2(u_{e\parallel}A_{1\parallel})\approx u_{e\parallel}\nabla_\perp^2A_{1\parallel}$.This form makes the interplay between the potentials and the charge densities more apparent and shows how the gyrokinetic corrections modify the standard Poisson equation. The term $ ( 4\pi\frac{\rho_{th}^{2}}{\lambda_{D}^{2}} +1 )$ acts as an effective dielectric coefficient modifying the Laplacian of the scalar potential, and the coupling to the vector potential $A_{1 \parallel}$ is mediated by the electron parallel flow velocity $u_{e \parallel}$.

\subsection{Parallel Ampere}

We now turn to the derivation of the parallel Ampere equation from the action integral. Here, we compute only those terms that depend on the magnetic potential. Employing the variational derivative approach, we first derive the homogeneous solution of the electromagnetic action

 \[
\frac{\delta}{\delta \textbf{A}_{1}(\textbf{x})}{\mathcal{S}}_{f}\circ\hat{\chi}_{1}(\textbf{x})= \\
\]

\begin{equation}
 -\frac{\delta}{\delta \textbf{A}_{1}}\int  \frac{1}{8\pi}\left|\nabla\times[\textbf{A}_{0}(\textbf{x})+\varepsilon_{\delta}\textbf{A}_{1}(\textbf{x})]\right|^{2}
\end{equation} 

Considering  all terms from the Lagrangian, we can derive the parallel Ampere equation 

\begin{equation*}
\frac{1}{4\pi}\nabla^{2}{A}_{1 \parallel}(\textbf{x})+\frac{u_{e\parallel}}{c}\frac{\rho_{e}^{2}}{\lambda_{D}^{2}}\nabla_{\perp}^{2}\phi_{1}(\textbf{x})+ \frac{\beta_{e}}{8\pi} \nabla_{\perp}^{2}{A}_{1\parallel}(\textbf{x})=
\end{equation*}

\begin{multline}
\frac{\varepsilon_{\delta}^{2}e^{2}}{mc^{2}}\int dW\bigl\langle F_{e}(\mathbf{x}-\rho)\bigr\rangle{A}_{1\parallel}(\textbf{x})  \\
-\varepsilon_{\delta}e \frac{1}{c}\int d\theta d\mu_{gy} v_{gy, \parallel} \left\langle F_{e}(\mathbf{x}-\mathbf{\mathbf{\rho}})\right\rangle \\
 +\sum_{i}\frac{q_{i}}{c}\int F_{i}(x,v)u_{i \parallel}d\Omega.
\end{multline}
Using a low wave number approximation for the gyroaveraged electron distribution,  and taking in consideration the complete electromagnetic potential we have

\begin{equation}
\begin{split}
\frac{c}{4\pi} \nabla_{\perp}^{2}{A}_{1\parallel} 
\left( 1 + \frac{\beta_{e}}{2} \right) 
&+ u_{e\parallel} \frac{\rho_{e}^{2}}{\lambda_{D}^{2}} \nabla_{\perp}^{2}\phi_{1}(\textbf{x}) \\
&= \frac{e^{2}}{m_e c} n_{e}{A}_{1 \parallel}(\textbf{x}) - I_{e} + \sum_i I_{i \parallel}.
\end{split}
\label{modified_ampere_eq}
\end{equation}

where $n_{i}=\int d\Omega F_{i}(x,v)$, and  $  \sum_{i}q_{i}\int F_{i}(x,v)u_{i \parallel}d\Omega=q_i\sum_{i}n_{i}u_{i \parallel} = \sum_{i } I_{i \parallel} $. We also considered $n_{e}=\int d\Omega F_{e}(\mathbf{x})$, and $I_{e}=e u_{e}n_e$. The first term on the right side is due to the velocity shift transformation performed in the gyrokinetic coordinates transformation detailed in \cite{Nathan}. This term is also associated with the magnetization current. Specifically for electrons, it indicates that a reduction in coordinates prevents the natural occurrence of diamagnetic current \cite{Frank-Kamenetskii1972}. It is important to note that, since we are currently solely focusing on perturbations perpendicular to the background magnetic field, we will defer the derivation of the perpendicular Ampere equation to future work.

\section{Electrostatic Linear Results:\protect\\ Waves and Dispersion Relation}\label{sec:level5}

Despite the immense complexity and the myriad of physical phenomena occurring in solar wind, within the framework of plasma physics, a valuable and insightful exercise for model validation involves conducting a wave analysis. Waves are associated with energy transfer, though not necessarily with dissipation, and frequently manifest in turbulent systems \citep{PhysRevResearch.2.043253, Plasma_Astro}. Here, we undertake a linear analysis to explore the waves inherent in our model. Thus, we adapted the solver FIDEL \citep{Felix}. We compare results with two other linear solvers: the fully kinetic DSHARK \citep{DSHARK} and the hybrid kinetic-ion/fluid-electron code HYDROS \citep{HYDROS}. We focus on three primary wave solutions of interest in space physics: ion Bernstein waves, the high-frequency limit case of the electrostatic lower hybrid wave, and ion acoustic waves.

\subsection{Ion Acoustic Waves}

Ion acoustic waves (IAW) are a longitudinal oscillation of charged particles in the plasma similar to acoustic waves on neutral gas.  IAW are subject to strong Landau damping. Ion acoustic–like fluctuations, possibly in the form of slow magnetosonic waves in magnetized plasmas, may contribute to dissipation mechanisms in the solar wind. They can also interact with the electromagnetic field. Such waves can be damped either collisionlessly via Landau damping or collisionaly through Coulomb interactions.

\begin{figure}
\begin{center}
\includegraphics[width=7.5cm]{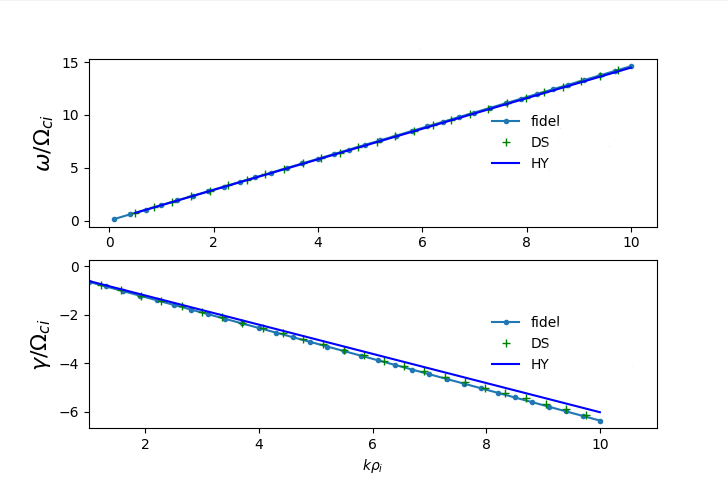}
\end{center}
\caption{Almost parallel ($\theta=5$) Ion Acoustic Wave frequency (top) and damping rate (bottom) for HYDROS (dark blue), DSHARK (green), and FIDEL (light blue).  }\label{IAW_t5}
\end{figure}

In the present work, we studied IAW with a recently modified version of FIDEL, where we also included the Maxwellian Laplacian in the Poisson equation, derived in section 5. The term referred to as the "Maxwellian Laplacian" is seen as the first term in equation \ref{final_ampere_eq}. It originates from the divergence of the electric field, a consequence of the charge density imbalance inherent in Poisson's equation. This modification allows us to better study high frequency waves, a point discussed in more detail in the next subsection. The linear dispersion relation $\omega \propto k $ characteristic of ion-acoustic waves is observed in Figure\ref{IAW_t5}. For sufficiently large wavenumber k, significant differences arise between the kinetic model and a fluid description with adiabatic electrons. In a purely fluid framework, ion-acoustic waves are undamped (neglecting explicit viscosity or resistivity), whereas in the kinetic treatment Landau damping leads to finite damping rates. The discrepancy therefore reflects the absence of collisionless phase mixing in the fluid closure.. 

For a plasma with ion temperature smaller than electron temperature, and assuming a large wavelength limit, the dispersion relation takes the form of

\begin{equation}
\left ( \frac{\omega}{k} \right ) ^2 =\frac{ c_{s}^{2}}{1+k^{2} \lambda_{D e}^{2}}+3 \frac{\kappa_B T_{s 0}}{m_{s}},\label{IAW_analytic}
\end{equation}

where $c_s$, $\lambda_{D e}$, and $\kappa_B $, represents sound speed, Debye length and Boltzmann constant, respectively. Details on the derivation for our model can be found elsewhere \cite{Nathan}. It is important to emphasize that IAW propagates with constant velocity, where the velocity lies between the thermal velocities of the species present on the medium. Our model also allows us to study parallel propagating IAW.

\begin{figure}
\begin{center}
\includegraphics[width=7.5cm]{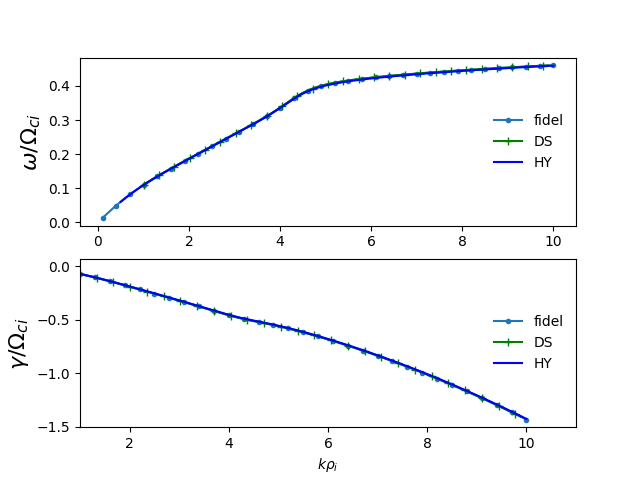}
\end{center}
\caption{Perpendicular ($\theta=85$) Ion Acoustic Wave frequency (top) and damping rate (bottom) for HYDROS (dark blue), DSHARK (green), and FIDEL (light blue).  }\label{IAW_t85}
\end{figure}

The Landau damping of IAW has been extensively discussed \cite{IAW_1}, where  Landau damping  strength is proportional to the ion thermal speed. The presence of IAW has also been associated with the increase of electron distribution anisotropy\cite{IAW_3}, and also solar wind heating\cite{IAW_4}.

\begin{figure}
\begin{center}
\includegraphics[width=6.5cm]{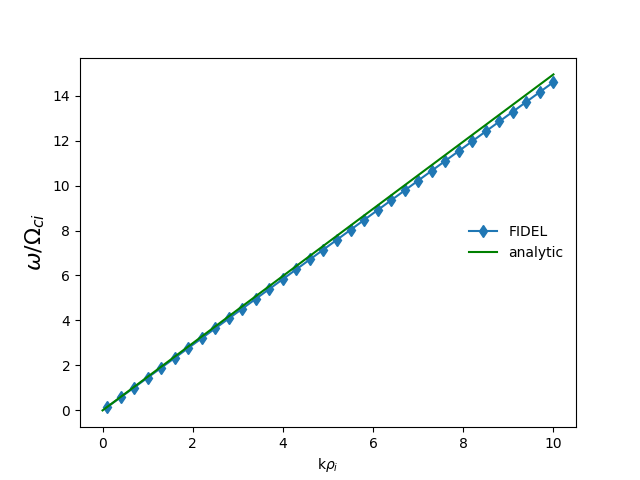}
\end{center}
\caption{Parallel ($\theta = 5$) comparison between the analytic IAW and FIDEL.}\label{IAW_analytic_fig}
\end{figure}

In the limit of large wave lengths, that is  $k^{2} \lambda_{D e}^{2} \ll 1$, the dispersion relation depends only on sound speed and temperature ratio, i.e.  

\begin{equation}
\omega^2 = c_{s}^2 k^2  \left ( 1 +3 \frac{T_i}{T_e} \right ).
\end{equation}
We also examined how our model aligns with the analytic solution (Figure \ref{IAW_analytic_fig}). For lower wavenumbers, the model reproduces previous FIDEL results. The inclusion of the Maxwellian Laplacian term improves the high-k behavior, extending the agreement with the analytic solution to shorter wavelengths. At sufficiently large k, however, discrepancies reappear, likely reflecting the limitations of the reduced model.

The presence of IAW modes in our model marks a significant milestone in creating a reduced framework. At a lower computational cost, it can describe key kinetic phenomena that may be crucial for understanding energy dissipation in solar wind.

\subsection{Ion Bernstein Waves}

\begin{figure}
\begin{center}
\includegraphics[width=7.5cm]{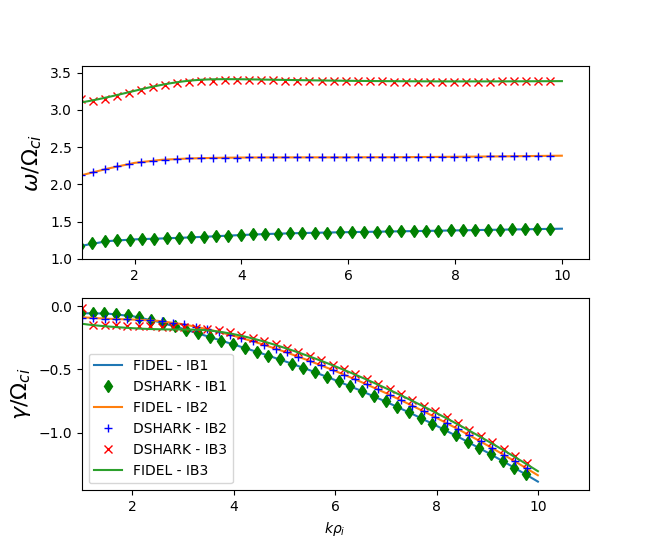}
\end{center}
\caption{Perpendicular ($\theta=85$) propagating Ion Bernstein Wave frequency (top) and damping rate (bottom) for different IBW modes}\label{IBW}
\end{figure}

Ion Bernstein waves (IBWs) are electrostatic, quasi-perpendicular waves in magnetized hot plasmas with frequencies near harmonics of the ion cyclotron frequency. Such waves interact with kinetic Alfven waves (KAW) \citep{IBW_2}, and are subject to strong localized electron Landau damping \citep{IBW_1}. Although IBWs arise in electromagnetic theory, our reduced model retains sufficient physics to capture the corresponding branch in the dispersion relation. In Figure \ref{IBW} we show a comparison between FIDEL and DSHARK simulations. For HYDROS, the numerical root-finding procedure exhibited convergence issues in this parameter regime, leading to unreliable solutions; these results were therefore excluded from the present comparison.

The general format of our analytic IBW dispersion relation looks like 

$$
k_{\perp}^{2}+2\left(\frac{\omega_{p}}{v_{t}}\right)^{2}e^{-\frac{1}{2}k_{\perp}^{2}\rho_{i}^{2}}\sum_{n=-\infty}^{\infty}I_{n}(\frac{1}{2}k_{\perp}^{2}\rho_{i}^{2})\left[1+\zeta_{0}Z\left(\zeta_{n}\right)\right]=0,
$$
where $I_n$ stands for modified Bessel function and Z stands for the plasma function. More details about the derivation can be found elsewhere \citep{Nathan}. 

\begin{figure}
\begin{center}
\includegraphics[width=7.5cm]{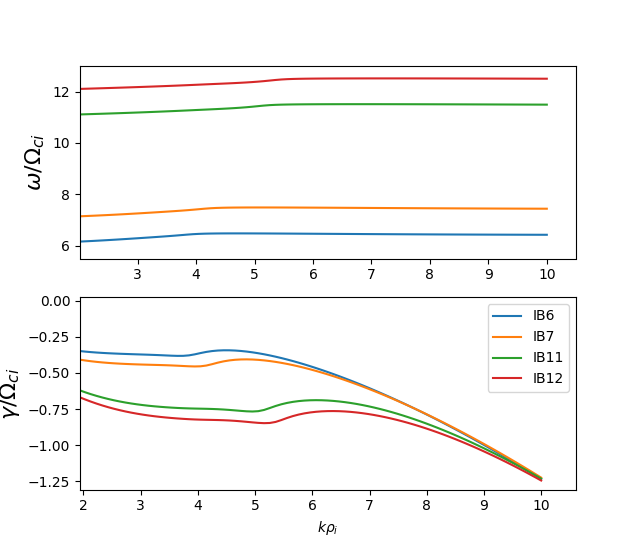}
\end{center}
\caption{Perpendicular ($\theta=85$) propagating Ion Bernstein Wave frequency (top) and damping rate (bottom) for higher IBW modes.  }\label{IBW_6}
\end{figure}

The agreement between both codes indicates that the reduced model is also capable of reproducing important fully kinetic linear features. It is imperative to have a model that can properly reproduce linear modes, as mode coupling between different wave solutions is believed to play a role in turbulence dissipation in collisionless plasmas. Furthermore, the presently modified FIDEL solver is also capable of reproducing higher IBW modes ( Figure\ref{IBW_6}), which is a  relevant point, also raised by \citep{IBW_2}.

The presence of higher IBW modes is likely due to the presence of the Maxwellian Laplacian in the Poisson equation. As discussed in more detail in following section, the use of the Maxwellian Laplacian allows for the computation of higher frequency waves in the system, which is important in the study of ion cyclotron range energy dissipation in solar wind turbulence. 

\subsection{Waves with $\omega \gg \Omega_{ci}$}\label{high_fre_sub}

Turbulent heating is thought to be one of the sources of ion heating in solar wind evolution \citep{high_freq_1, high_freq_2, high_freq_3}. Magnetic reconnection, Landau damping and transit time damping are thought to be the main processes involved in the dissipation mechanism at kinetic scales \citep{high_freq_4, high_freq_5}. The use of gyrokinetics to study solar wind turbulence has been questioned due to the $\omega \ll \Omega_i$ restriction described in Ref. \citep{IBW_2}. In the present model, we demonstrate that the drift kinetic limit of our hybrid kinetic-gyrokinetic formulation contains the lower hybrid wave, demonstrating that the present formulation extends beyond the standard gyrokinetic ordering and captures higher-frequency branches such as the lower-hybrid mode, which may provide a useful extension toward simulations that bridge standard gyrokinetic KAW physics and higher-frequency dynamics.

When looking at high frequency waves, our model's dispersion relation takes the form 

\begin{equation}
\omega^{2}=\frac{1}{2}\left(\Omega_{ci}^{2}+\frac{\omega_{pe}^{2}k_{\parallel}^{2}+\omega_{pi}^{2}k^{2}}{c_{v}k^{2}+\frac{\omega_{pe}^{2}}{\Omega_{ce}^{2}{}_{\perp}}k_{\perp}^{2}}\right) \label{high_freq_eq}
\end{equation}
\[
+\frac{1}{2}\sqrt{\left(\Omega_{ci}^{2}+\frac{\omega_{pe}^{2}k_{\parallel}^{2}+\omega_{pi}^{2}k^{2}}{c_{v}k^{2}+\frac{\omega_{pe}^{2}}{\Omega_{ce}^{2}{}_{}}k_{\perp}^{2}}\right)^{2}+\frac{4(\omega_{pe}^{2}+\omega_{pi}^{2})\Omega_{ci}^{2}}{c_{v}k^{2}+\frac{\omega_{pe}^{2}}{\Omega_{ce}^{2}{}_{}}k_{\perp}^{2}}k_{\parallel}^{2}}.
\]

In this context, solutions  that satisfy $\omega \gg\Omega_{i}$ are compared with their analytical counterpart to  examine whether the present model is capable of describing high frequency phenomena. To understand high frequency waves at various directions to the background magnetic field, we performed a $\theta$ scan. \\

\begin{figure}
\begin{center}
\includegraphics[width=8.5cm]{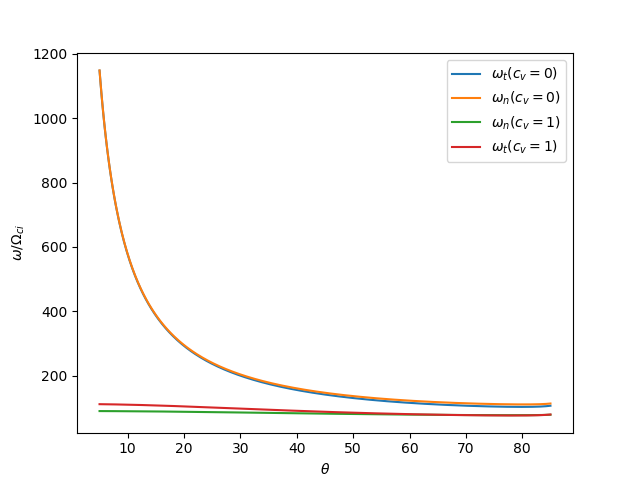}
\end{center}
\caption{$\theta$ scan at $k\rho_i=10$ and with $\frac{T_i}{T_e}=1$ of $\omega \gg \Omega_{ci}$.}\label{high_freq_t}
\end{figure}

Figure \ref{high_freq_t} shows the high frequency wave of Eq.(\ref{high_freq_eq})with ($c_v=1$) and without ($c_v=0$) the Maxwellian Laplacian,  at $k\rho_i=10$ for various propagating angles. At small $\theta$ (quasi-parallel propagation), the case without the Maxwellian Laplacian exhibits a spurious blow-up of the frequency, reflecting the limitations of the reduced drift-kinetic ordering$ (\omega \ll \Omega_i)$. The effect of the Maxwellian Laplacian is seen on the numerical ($\omega_n$, green) and theoretical ($\omega_t$, red) curves (Figure \ref{high_freq_t}). Nevertheless, for small propagation angles $(\theta \rightarrow 0$, corresponding to large $k_\parallel$ we observe that the effect of the Maxwellian Laplacian supersedes that of the electron polarization response. For nearly parallel propagation, the frequency approaches the electron plasma frequency, consistent with Langmuir-like oscillations. Assuming representative solar wind parameters 3 $ cm^{-3}$, and the background magnetic field is around $10^{-5}$ Gauss \cite{Verscharen}, we obtain $\frac{\omega_{pe}}{\Omega_{ci}} \sim 90$. This serves only as an ordero-if-magnitude estimate for comparison with in-situ conditions. Although the model is not designed as a full high-frequency electromagnetic framework, it retains sufficient physics to capture wave behavior approaching the Langmuir frequency in the $k \approx k_\parallel$ limit. For small values of $\theta$, both the numerical and theoretical linear limits of the model (Figure \ref{high_freq_t}) predict values of the order of magnitude of the value computed using in situ data.

\section{Nonlinear electrostatic results: Landau Damping}\label{sec:level6}

Recent progress has underscored the significance of kinetic-scale effects, which critically influence the dynamics of reconnection processes\citep{Chen2019, PSP_9}. To accurately capture these effects, particularly at electron scales, kinetic models such as the hybrid gyrokinetic-electron and fully kinetic-ion formulations could be of great importance. In this section, we present simulation results obtained from our novel variational kinetic approach.

\subsection{SSV: Quick Numerical Overview}

The Super Simple Vlasov (ssV) code, solves fully kinetic ion physics alongside drift-kinetic electron physics within slab geometry employing periodic boundary conditions in the spatial dimension. It utilizes domain decomposition \citep{domain_decomposition} coupled with ghost-cell communication to efficiently solve high-dimensional Vlasov-Maxwell systems on multicore architectures, partitioning the simulation domain into independent subdomains assigned to separate processor cores with optimized boundary exchanges to ensure accuracy and scalability. Localized numerical schemes reduce inter-core communication, while kinetic boundary exchange protocols maintain consistency in charge and current densities without additional corner-based exchanges. Notably, the electromagnetic field solver operates entirely in Fourier space, which enhances numerical stability and reduces parallelization overhead in large-scale simulations.

The code applies a Strang splitting algorithm to efficiently solve the high-dimensional Vlasov equation by sequentially evolving the distribution function separately in position space and velocity space, preserving second-order temporal accuracy. To manage non-commuting advection operations in velocity space, it utilizes cascade interpolation combined with back-substitution, along with a Boris push discretization scheme. This approach significantly reduces computational complexity while maintaining numerical precision.

The code employs a semi-Lagrangian method \citep{semi_lagrange}, integrating fixed Eulerian grid structures with Lagrangian trajectory tracing to accurately solve electron advection equations without restrictive Courant-Friedrichs-Lewy (CFL) conditions. This hybrid method computes backward trajectories using an iterative Newton method and interpolates the electron distribution function through third- or fifth-order Lagrange polynomials, significantly reducing numerical diffusion while maintaining computational efficiency and second-order accuracy. Additionally, the code incorporates a semi-Lagrangian monotonicity- and positivity-preserving (SLMPP) framework \citep{slmp}, specifically using the fifth-order SLMP5 scheme, which enables high-order accurate simulations of collisionless astrophysical plasmas, effectively handling sharp gradients and multiscale dynamics.

For a more detailed overview of the numerical aspects of the code, we refer the reader to  ~\citep{thatikonda2025}.

\subsection{SSV: Verification with nonlinear Landau damping test case}
As a nonlinear benchmark, we consider the nonlinear Landau damping problem, which is widely used to test the accuracy and conservation properties of Vlasov solvers \cite{cheng1976,sonnendrucker1999,filbet2001,crouseilles2010,liu2023}. The initial condition is given by a perturbed Maxwellian distribution,
\begin{equation}
f(0,x,v) = \frac{1}{\sqrt{2\pi}} \exp\!\left(-\frac{v^2}{2}\right)\bigl( 1 + \alpha \cos(kx) \bigr),
\label{eq:init_f}
\end{equation}
with perturbation amplitude $\alpha = 0.5$, wavenumber $k=0.5$, and periodic domain $x \in (0,L)$, $L = 4\pi$. The velocity domain is $v \in (-6,6)$.  

The simulations are performed with the \texttt{super simple Vlasov (ssV)} solver~\citep{thatikonda2025}. For the present study we restrict to the electrostatic branch and employ the semi-Lagrangian monotonicity-preserving fifth-order scheme (SLMP5). We use a timestep $\Delta t = 0.01$ and three different velocity-space resolutions, $(N_x,N_v)=(32,64)$, $(32,128)$, and $(32,256)$, to investigate resolution effects.

\begin{figure}[h]
    \centering
    \includegraphics[width=0.5\textwidth]{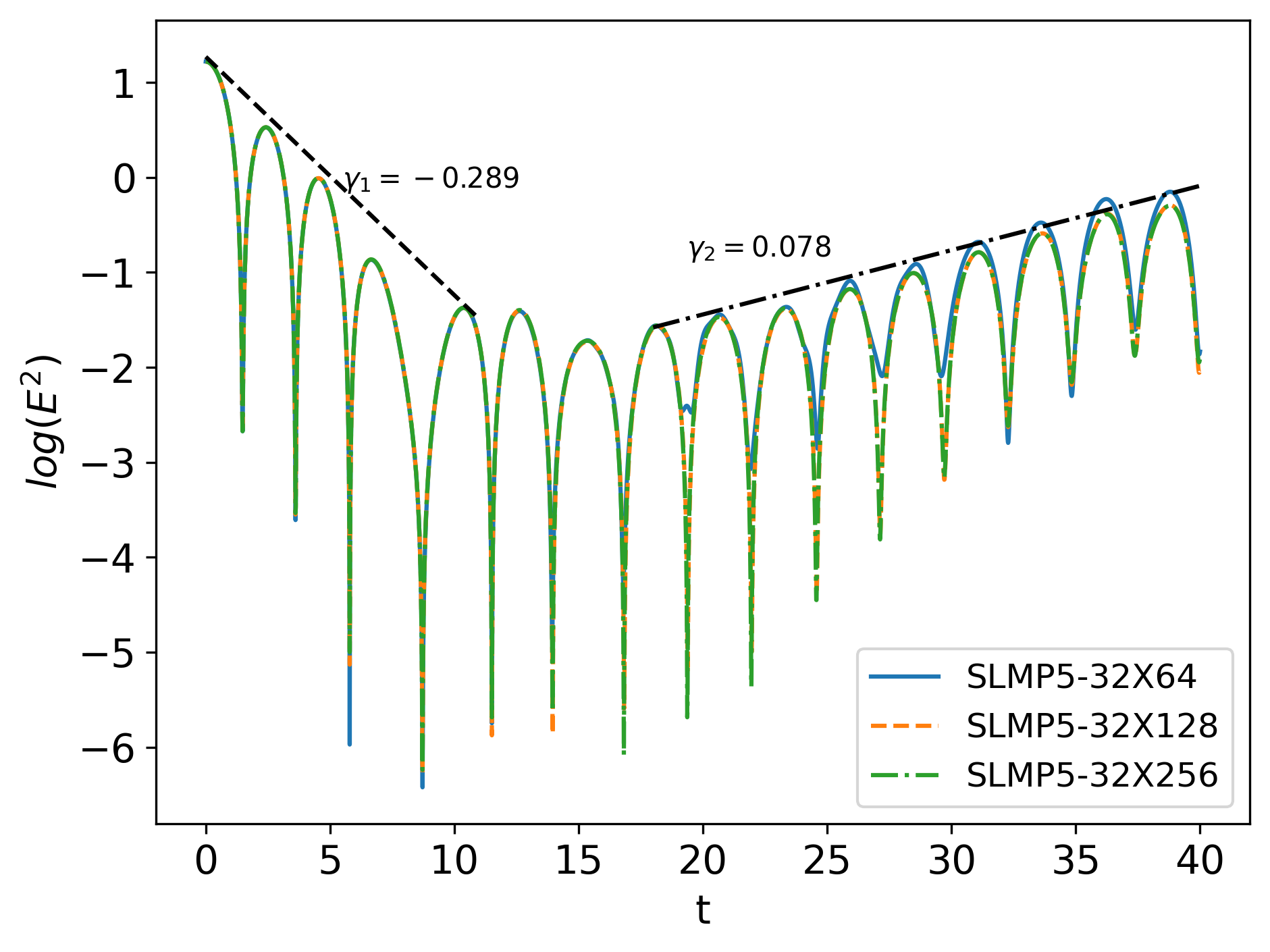} 
    \caption{Time evolution of the electric field energy $E^2(t)$ in logarithmic scale for
  the SLMP5 scheme at resolutions $(N_x,N_v)=(32,64)$, $(32,128)$, and $(32,256)$.
  Reference slope lines for the linear and nonlinear phases are overlaid.}
  \label{fig:E2_log_all}
\end{figure}

\begin{figure}[h]
    \centering
    \includegraphics[width=0.5\textwidth]{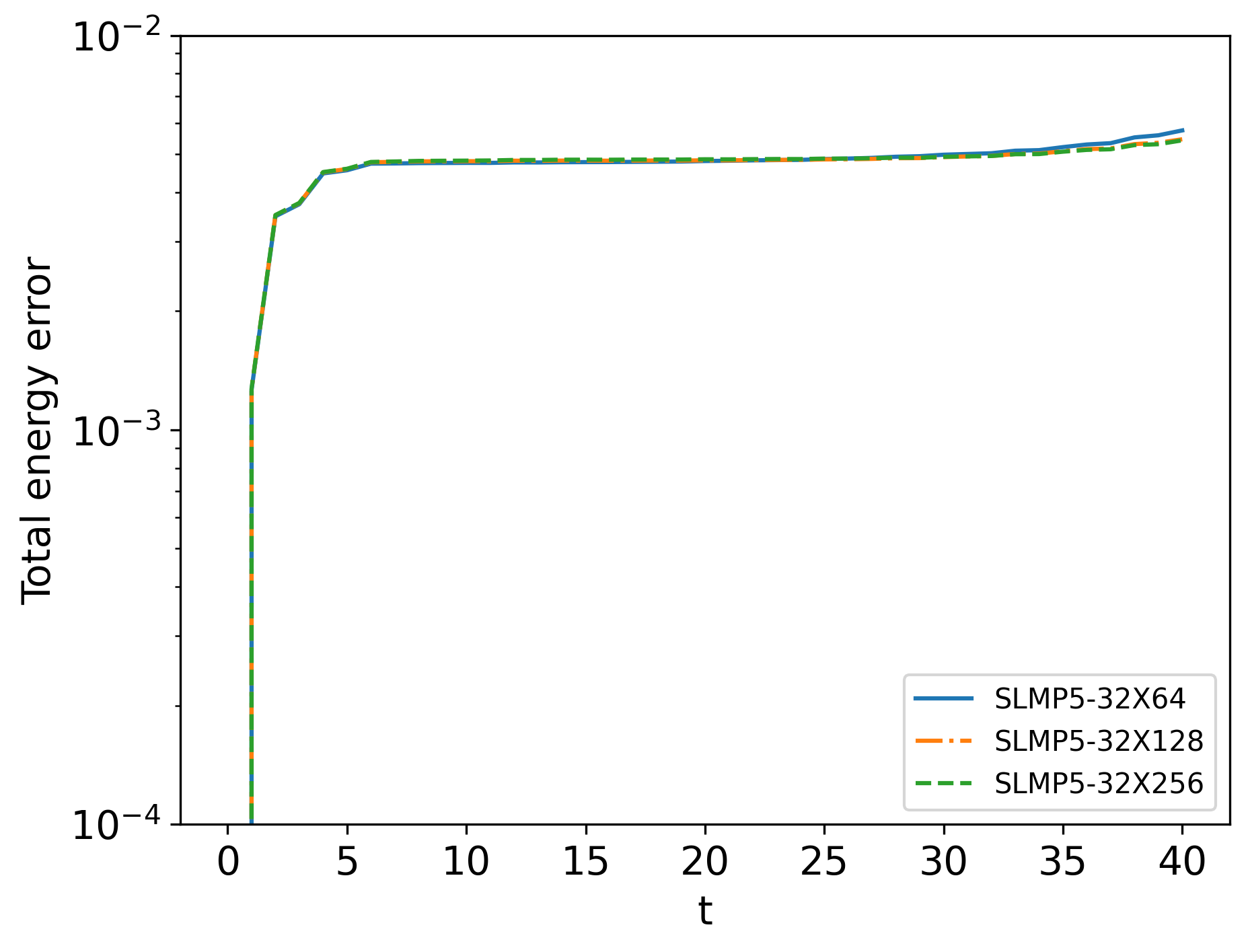} 
    \caption{Relative total energy error versus time for the SLMP5 scheme at
  $(N_x,N_v)=(32,64)$, $(32,128)$, and $(32,256)$.}
  \label{fig:energy_relerr}
\end{figure}

\begin{figure*}[t]
  \centering
  \begin{subcaptionblock}{0.32\textwidth}
    \centering
    \includegraphics[width=\linewidth]{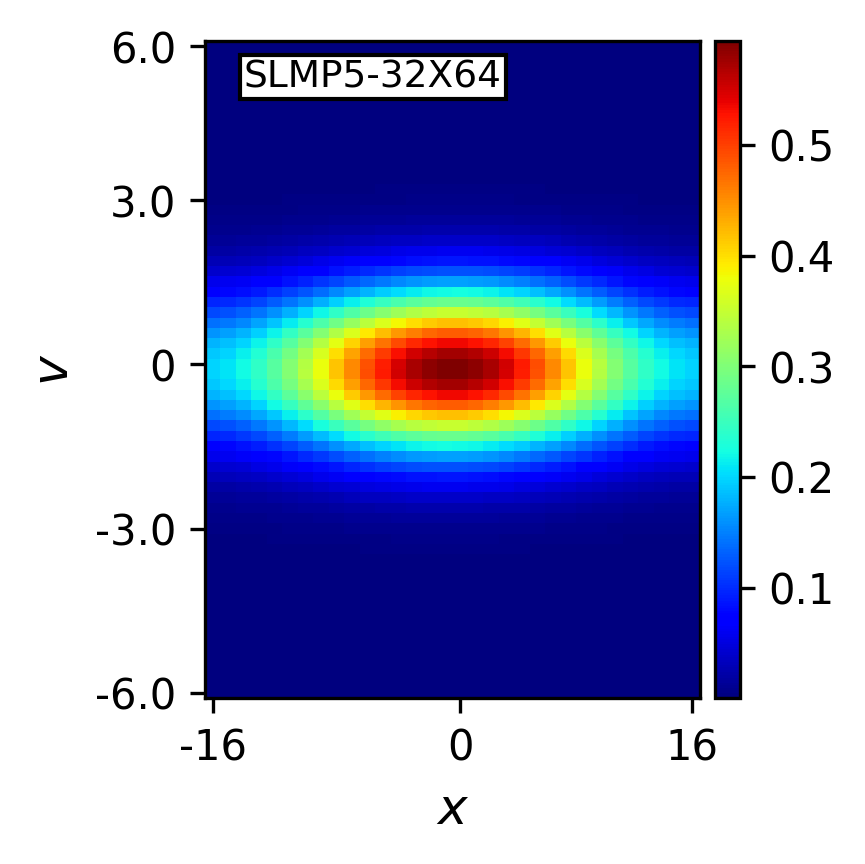}
    \caption{$32\times64$, $t=0$}
    \label{fig:PS_32x64_t0}
  \end{subcaptionblock}
  \begin{subcaptionblock}{0.32\textwidth}
    \centering
    \includegraphics[width=\linewidth]{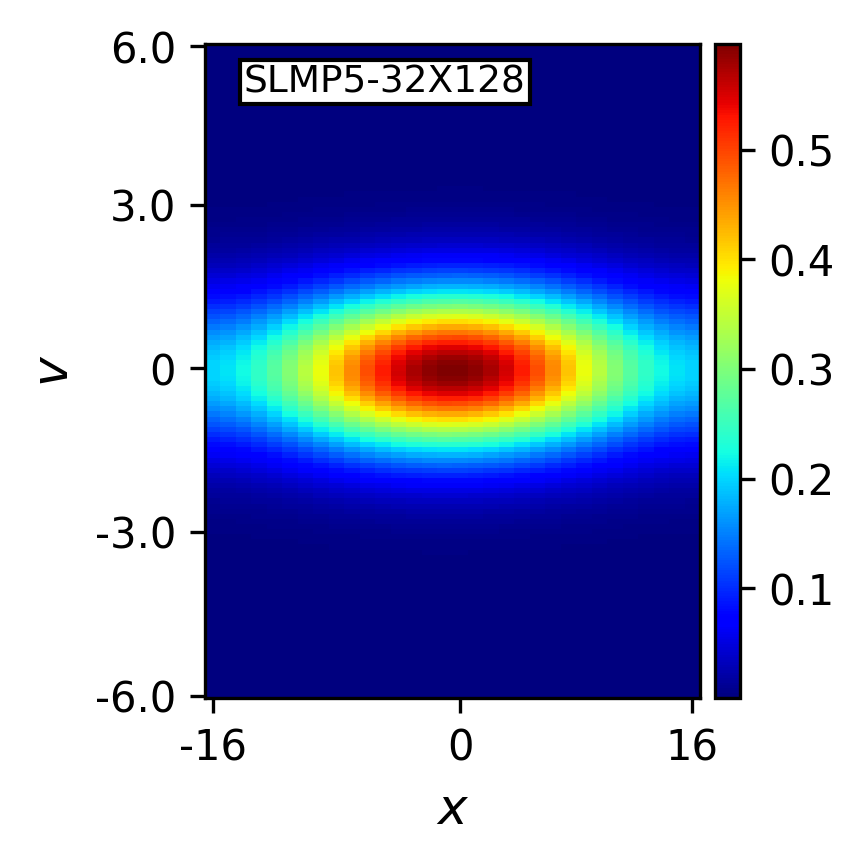}
    \caption{$32\times128$, $t=0$}
    \label{fig:PS_32x128_t0}
  \end{subcaptionblock}
  \begin{subcaptionblock}{0.32\textwidth}
    \centering
    \includegraphics[width=\linewidth]{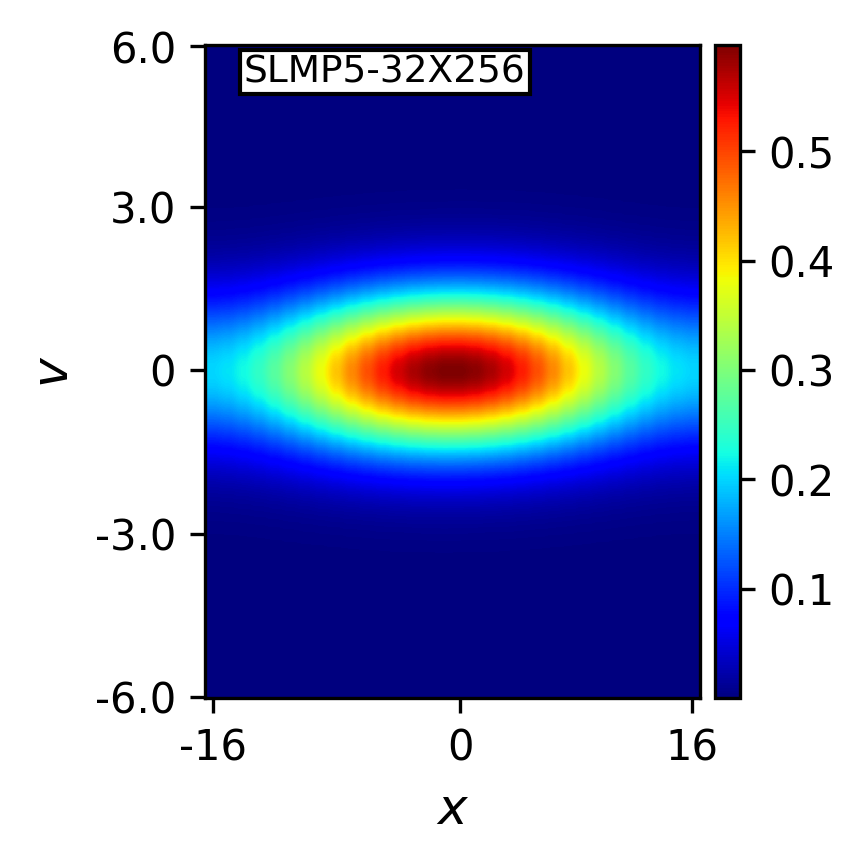}
    \caption{$32\times256$, $t=0$}
    \label{fig:PS_32x256_t0}
  \end{subcaptionblock}

  \vspace{0.6em}
  \begin{subcaptionblock}{0.32\textwidth}
    \centering
    \includegraphics[width=\linewidth]{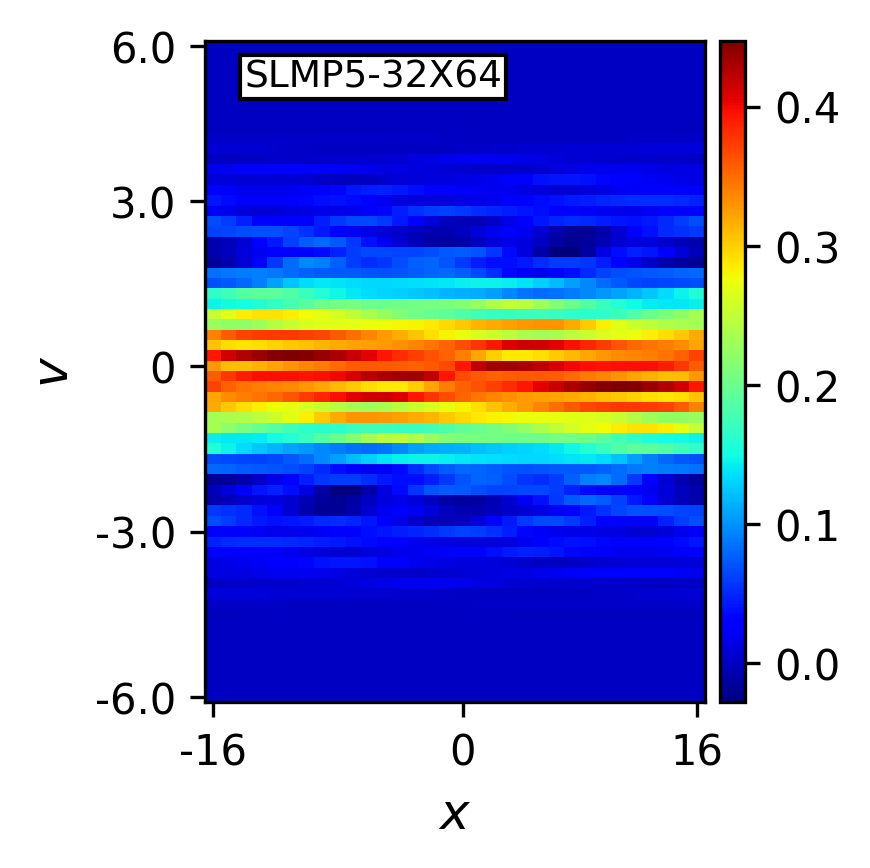}
    \caption{$32\times64$, $t=30$}
    \label{fig:PS_32x64_t30}
  \end{subcaptionblock}
  \begin{subcaptionblock}{0.32\textwidth}
    \centering
    \includegraphics[width=\linewidth]{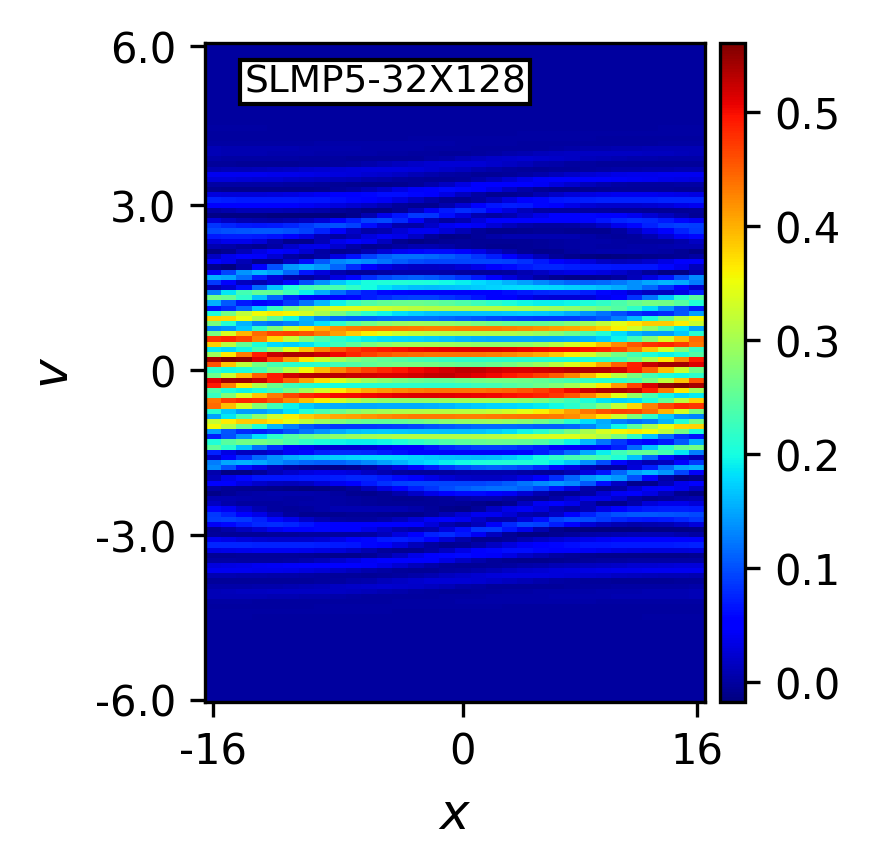}
    \caption{$32\times128$, $t=30$}
    \label{fig:PS_32x128_t30}
  \end{subcaptionblock}
  \begin{subcaptionblock}{0.32\textwidth}
    \centering
    \includegraphics[width=\linewidth]{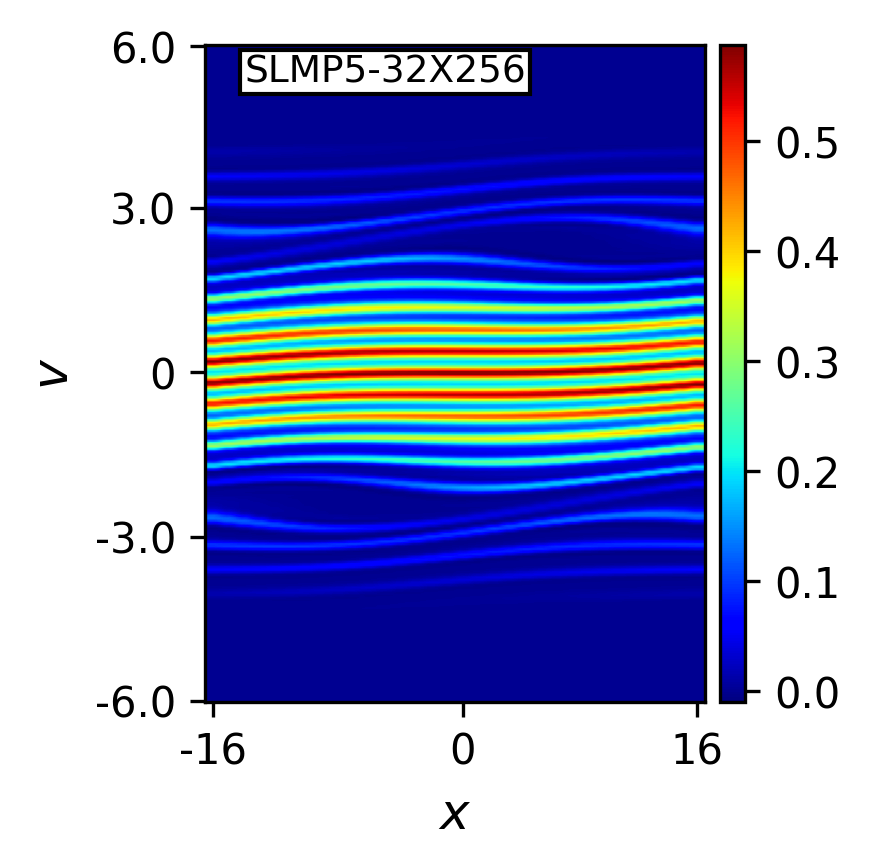}
    \caption{$32\times256$, $t=30$}
    \label{fig:PS_32x256_t30}
  \end{subcaptionblock}

  \caption{Phase-space distribution $f(x,v,t)$ at $t=0$ and $t=30$ for three grids (SLMP5).}
  \label{fig:phasespace}
\end{figure*}

\begin{figure*}[t]
  \centering
  \begin{subcaptionblock}{0.32\textwidth}
    \centering
    \includegraphics[width=\linewidth]{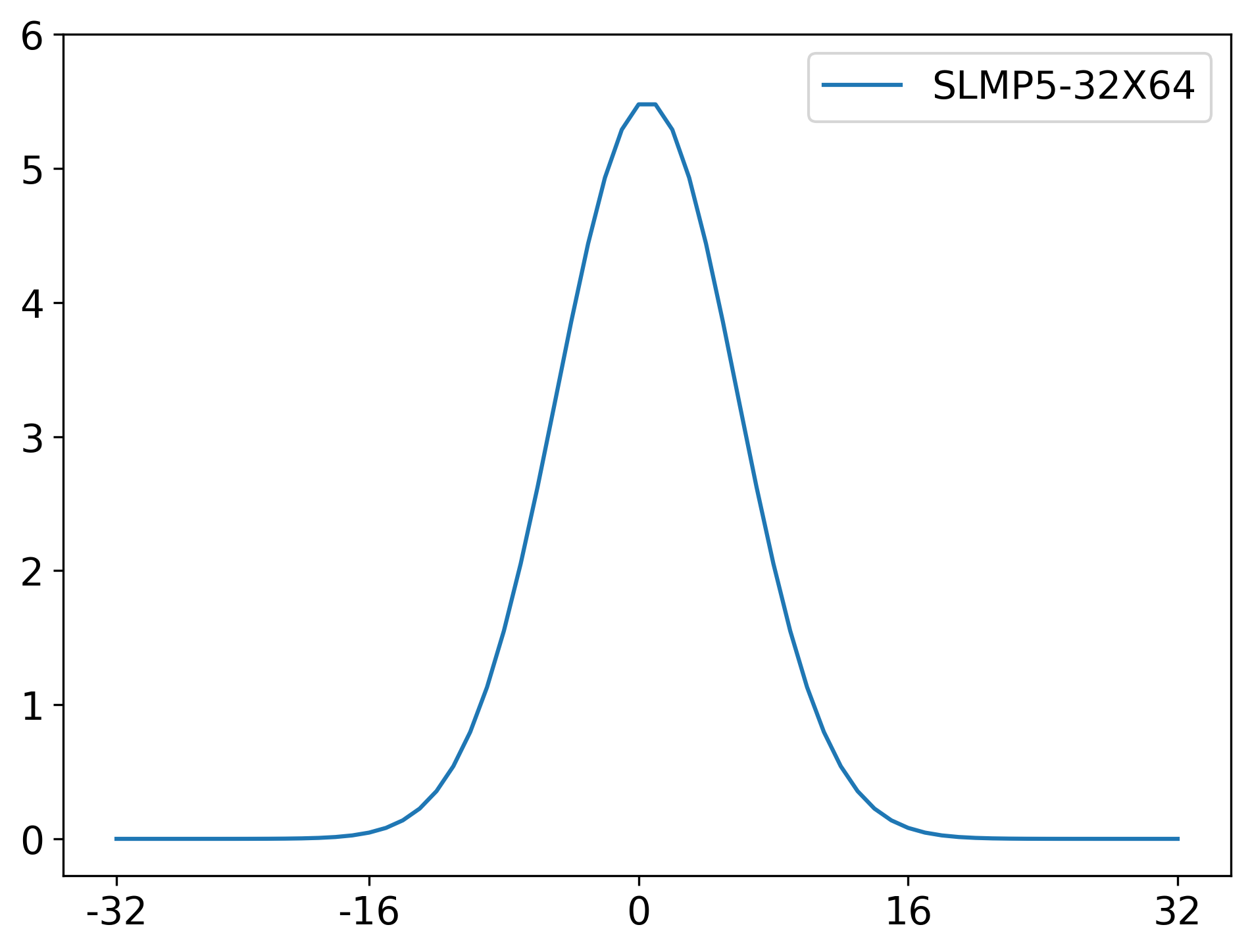}
    \caption{$32\times64$, $t=0$}
    \label{fig:intf_32x64_t0}
  \end{subcaptionblock}
  \begin{subcaptionblock}{0.32\textwidth}
    \centering
    \includegraphics[width=\linewidth]{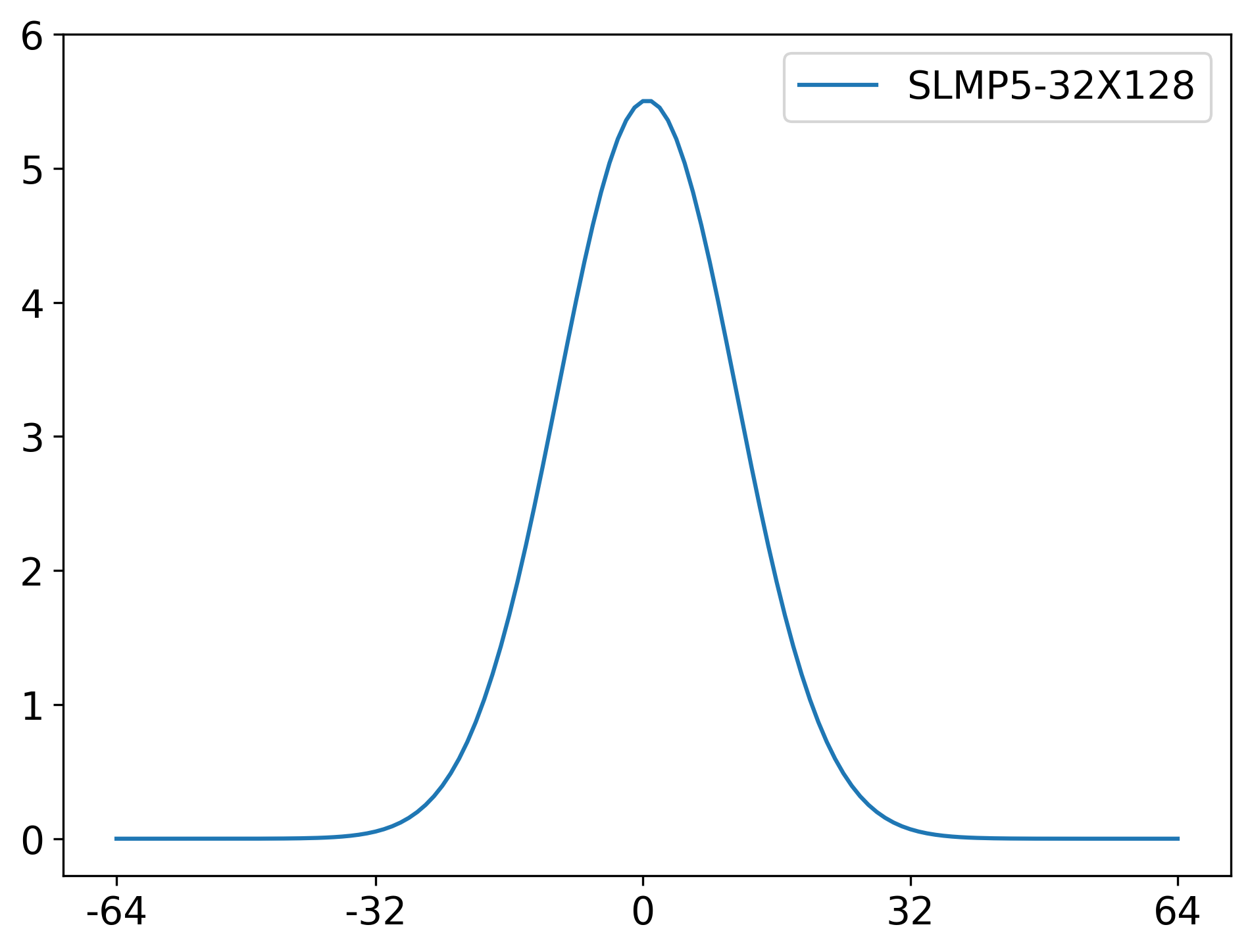}
    \caption{$32\times128$, $t=0$}
    \label{fig:intf_32x128_t0}
  \end{subcaptionblock}
  \begin{subcaptionblock}{0.32\textwidth}
    \centering
    \includegraphics[width=\linewidth]{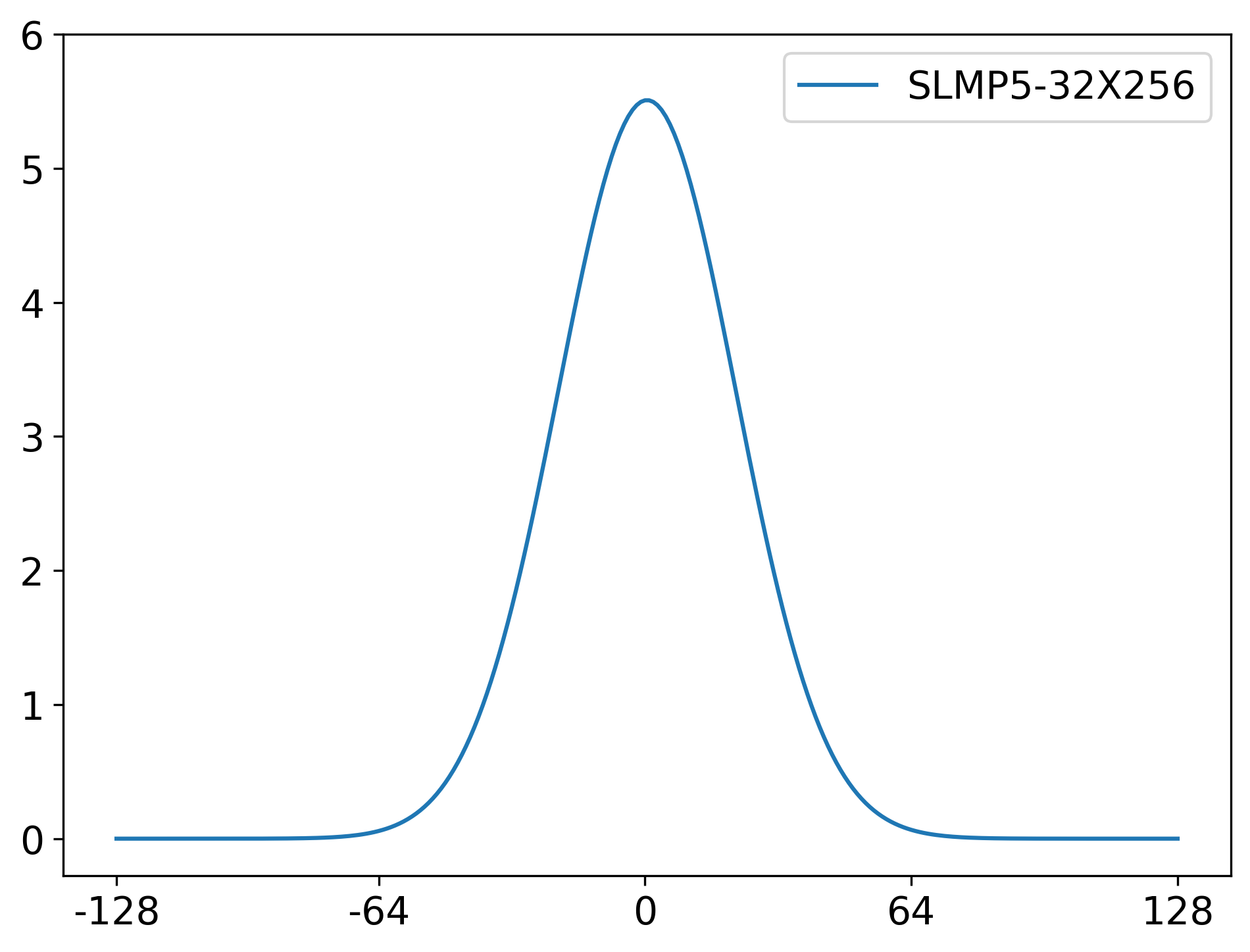}
    \caption{$32\times256$, $t=0$}
    \label{fig:intf_32x256_t0}
  \end{subcaptionblock}

  \vspace{0.6em}
  \begin{subcaptionblock}{0.32\textwidth}
    \centering
    \includegraphics[width=\linewidth]{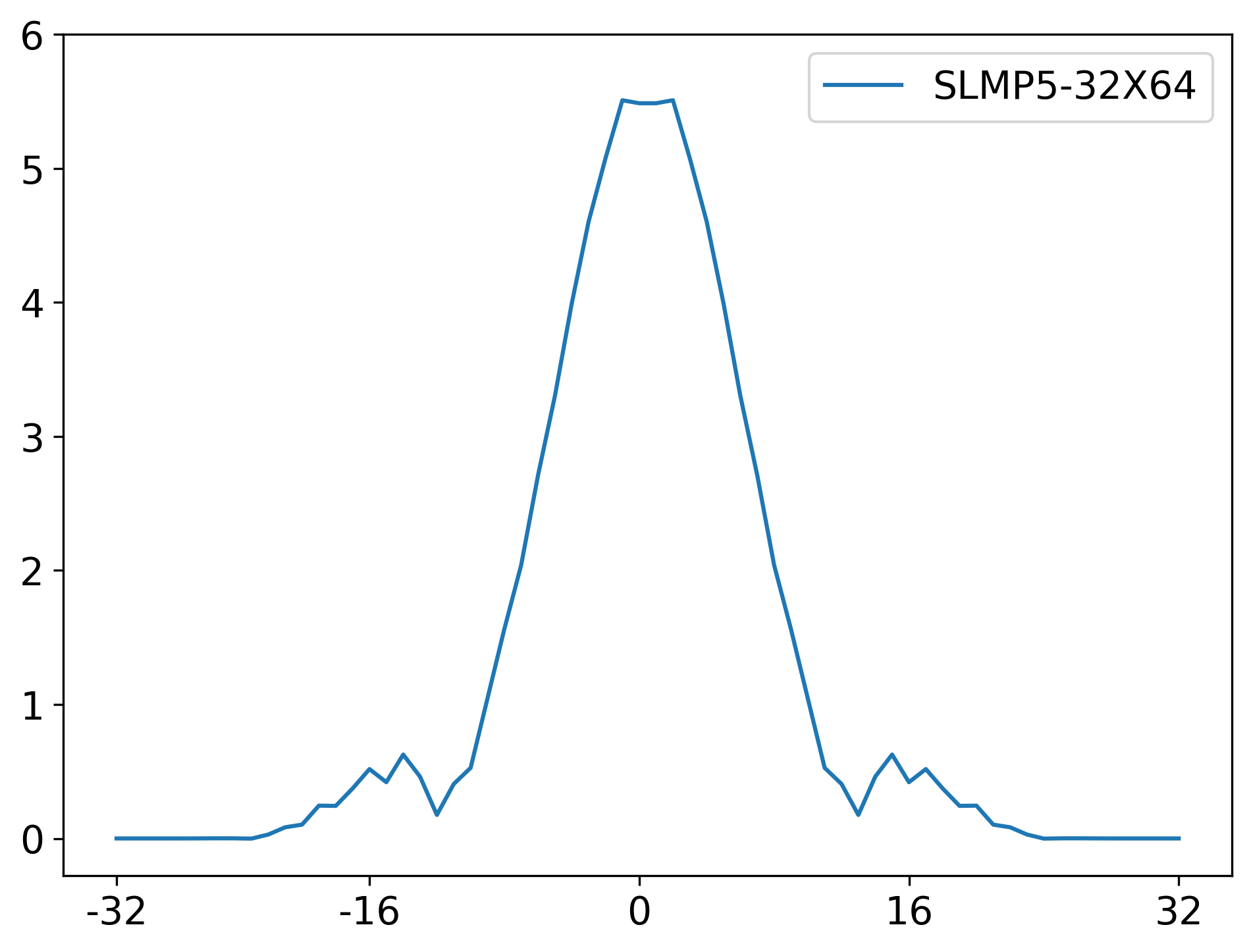}
    \caption{$32\times64$, $t=30$}
    \label{fig:intf_32x64_t30}
  \end{subcaptionblock}
  \begin{subcaptionblock}{0.32\textwidth}
    \centering
    \includegraphics[width=\linewidth]{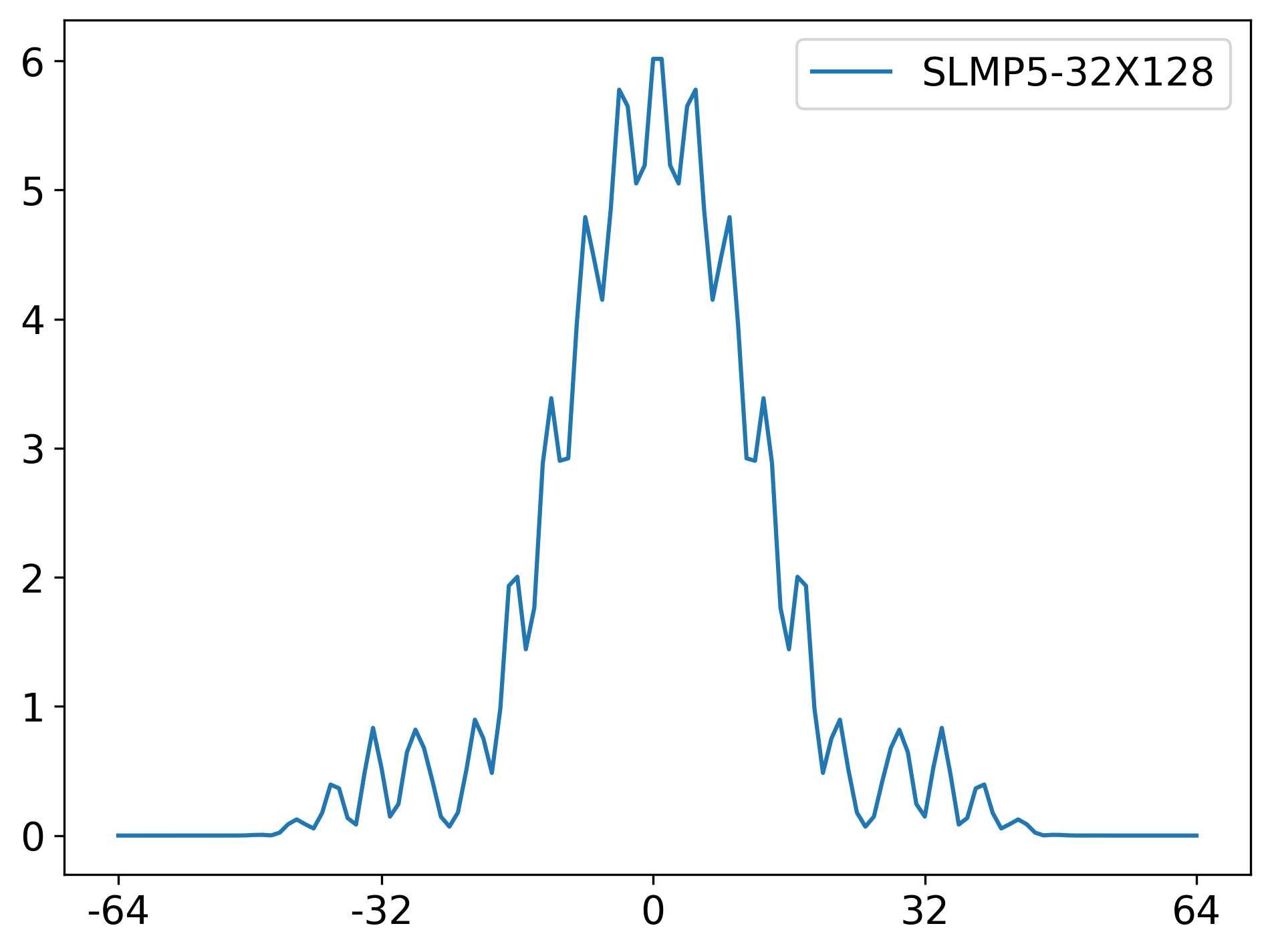}
    \caption{$32\times128$, $t=30$}
    \label{fig:intf_32x128_t30}
  \end{subcaptionblock}
  \begin{subcaptionblock}{0.32\textwidth}
    \centering
    \includegraphics[width=\linewidth]{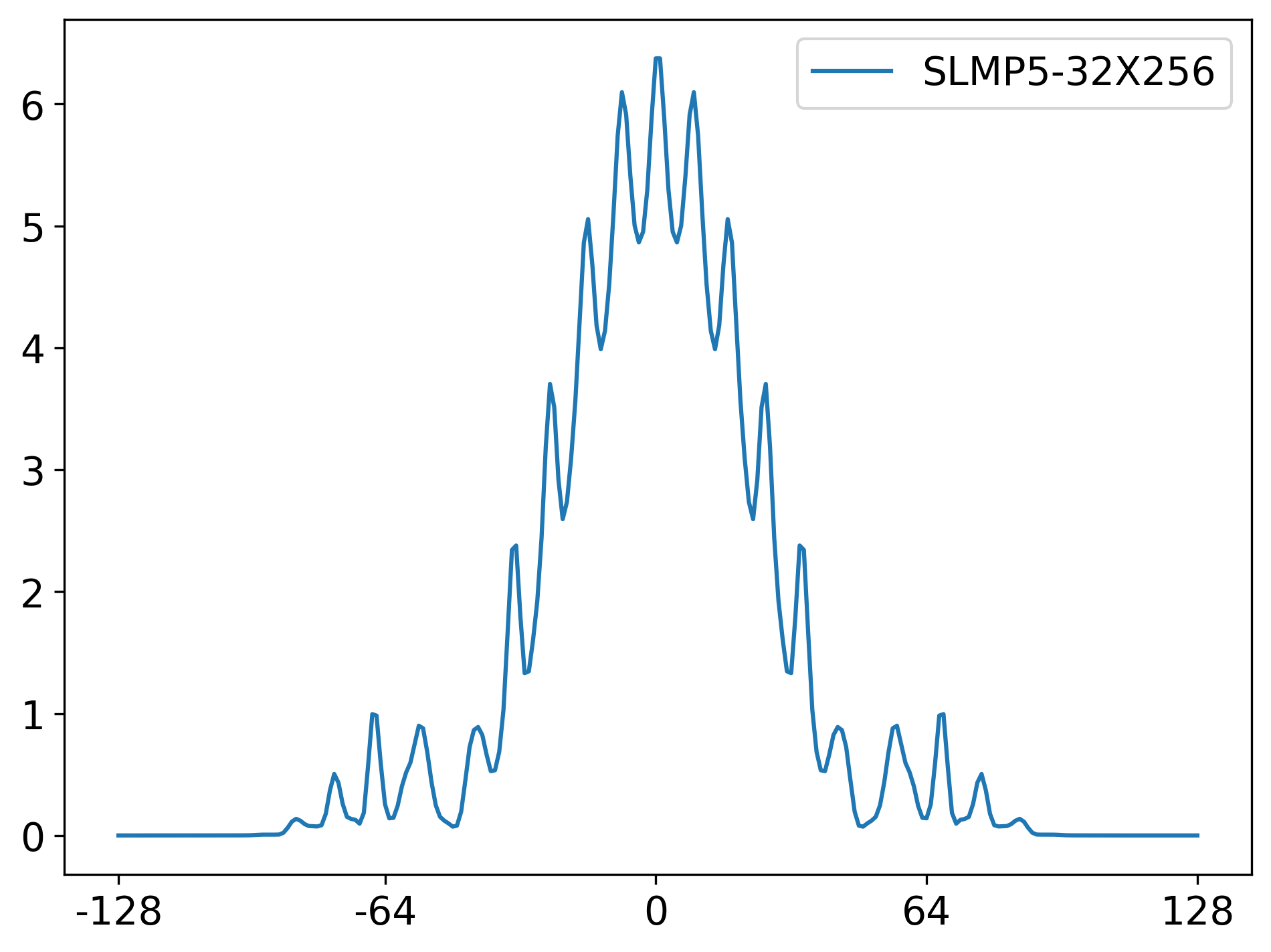}
    \caption{$32\times256$, $t=30$}
    \label{fig:intf_32x256_t30}
  \end{subcaptionblock}

  \caption{Spatially integrated distribution $\int f(x,v,t)\,dx$ at $t=0$ (top) and $t=30$ (bottom) for three grids (SLMP5).}
  \label{fig:intf_panels}
\end{figure*}

\paragraph*{Electric field energy evolution.}  
Figure~\ref{fig:E2_log_all} shows the time evolution of the electric field energy $E^2(t)$ in logarithmic scale. Two distinct regimes are observed: the linear damping phase with rate $\gamma_1 \approx -0.289$, followed by a nonlinear growth phase with $\gamma_2 \approx 0.078$. These values are consistent with theoretical predictions and previous numerical studies \cite{cheng1976,filbet2001,crouseilles2010,liu2023}, indicating that the SLMP5 scheme captures both the linear and nonlinear dynamics of Landau damping.
\paragraph*{Total energy conservation.}  
The total energy error, shown in Figure~\ref{fig:energy_relerr}, remains bounded for all resolutions without secular drift. This demonstrates that the discretization maintains the conservation properties of the system over long simulation times, which is essential for resolving nonlinear dynamics. Similar behavior has been reported in previous benchmark studies, such as Liu et al.~\cite{liu2023} and Chang et al.~\cite{chang2021}, and our results are in line with these findings across all grid resolutions considered here.
\paragraph*{Phase-space dynamics.}  
Phase-space contour plots of the distribution function at $t=0$ and $t=30$ (Figure~\ref{fig:phasespace}) highlight the formation of trapping and filamentation structures during the nonlinear stage. While all grids capture the qualitative dynamics, finer velocity resolutions provide sharper gradients and better preservation of fine-scale phase-space structures, reducing numerical diffusion.
\paragraph*{Spatially integrated distribution.}  
Figure~\ref{fig:intf_panels} presents the time evolution of the spatially integrated distribution function for the three grids. The results show convergence toward a consistent nonlinear plateau, with higher $N_v$ yielding a more accurate representation of the phase-mixed state.
\medskip

In summary, the nonlinear Landau damping test confirms that the \texttt{ssV} solver with SLMP5 reproduces the correct linear damping and nonlinear saturation, while maintaining good conservation properties. The observed agreement with benchmarks from the literature \cite{cheng1976,sonnendrucker1999,filbet2001,crouseilles2010,liu2023} supports the reliability of the solver for nonlinear plasma dynamics.
\section{Conclusion:\protect\\The efficacy of reduced models in space and astrophysics turbulence}\label{sec:level6}

Energy dissipation in solar wind  has been regarded as one of the outstanding unsolved problems in space plasma physics \citep{Goldstein2001}. While the full understanding of the various mechanisms that take place on the solar corona and solar winds are far from conclusive, it is expected that turbulence plays a major role \citep{9_outstanding_problems}. The multi-scale nature of turbulence in space and astrophysics plasma imposes a considerable barrier in computing those systems from the injection to the dissipation range scale. To bridge this gap, reduced models are often used, and despite their constraints, a good agreement has been shown \citep{Howes2006,Told}. Despite this, to better understand energy dissipation at ion scale, one needs to take into account constraints that violate the gyrokinetic conditions \citep{IBW_2}.  

With this in mind, we have derived a hybrid model that takes into consideration a reduced description for electrons and ions in their fully kinetic nature, to better compute ion scale range phenomena. We have seen how the reduced dynamics can be coherently coupled with the fully kinetic description through a coordinate transformation on the field equations frame of reference. Linearly, we have benchmarked our system with two dispersion solvers, HYDROS and DSHARK. We have seen that our system reproduces well not only IAW and IBW, but also higher frequency waves. A discussion about the relationship between IBW and high frequency waves points to the direction that reduced hybrid drift-kinetic electrons and fully kinetic ions are already capable of depicting higher frequency phenomena and, therefore, could be considered a step towards a reduced model capable of describing energy dissipation in solar wind. 

Nevertheless, the next steps are fundamental to cement the present results. The next stage involves implementing an electromagnetic dispersion relation solver. This will allow us to study KAW and their coupling with IBW, and, when cross-analysed with Landau damping \citep{IBW_2, PhysRevResearch.2.043253} and ion Cyclotron damping \citep{Chandran2010}, could shed some light on the understanding of one of the many dissipation mechanisms taking place in solar wind turbulence. Moreover, the nonlinear implementation of the system is already ongoing and could further provide support to the model's validity .

\appendix

\section{Dispersion Relation}

To perform a preliminary numerical analysis of our system of equations, we look at a simpler case. Here, we construct the Vlasov equation using the equations of motion derived in section 4 above, and we perform a drift kinetic and electrostatic approximation to the electron species of the system. The equation reads
 \begin{equation}
     \frac{\partial F}{\partial t} + \dot{X}_{gy} \boldsymbol{\cdot} \nabla_{gy} F + \dot{v}_{gy, \parallel} \partial_{\bar{v}_{gy, \parallel}} F = 0,
 \end{equation}
 
that is

\begin{equation*}
   \frac{\partial F}{\partial t} +  \frac{\textbf{B}^{*}}{mB_{\parallel}^{*}} \left \{ m {v}_{gy, \parallel} + \varepsilon_\delta \frac{e}{c} \left < A_{1 \parallel }\right > \right \} \boldsymbol{\cdot} \nabla_{gy} F
\end{equation*}
  
\begin{equation*}
     + \frac{c \hat{b}}{e B^*_{\parallel}} \times \left ( \mu_{gy} \nabla_{gy} B (X_{gy}) - \varepsilon_\delta e \nabla \left < \phi_1 \right > +\varepsilon_\delta \frac{e}{c}{v}_{gy, \parallel} \nabla \left < A_{1 \parallel }\right >  \right ) \boldsymbol{\cdot} \nabla_{gy} F
\end{equation*}
      
\begin{equation}\label{25}
\begin{aligned}
- \frac{\mathbf{B}^{*}}{m B_{\parallel}^{*}}
\boldsymbol{\cdot}
\Big( \mu_{gy} \nabla_{gy} B(X_{gy})
- e \varepsilon_\delta \nabla \langle \phi_1 \rangle \\
\quad
+ \frac{e}{c} \bar{v}_{gy,\parallel} \varepsilon_\delta \nabla \langle A_{1\parallel} \rangle
\Big) \partial_{v_{gy,\parallel}} F
= 0
\end{aligned}
\end{equation}
      
In the first line, we have the term representing the parallel motion. The first term in the second line represents the grad-B drift, while the second term denotes the E$\times$B drift. The last term in that line represents the B flutter. The final line corresponds to parallel acceleration.

First, we consider that $A_1$ vanishes, the Vlasov equation becomes then
 
\begin{equation}
\begin{split}
\frac{\partial F}{\partial t} 
&+ \frac{\textbf{B}^{*}}{m B_{\parallel}^{*}} m {v}_{gy, \parallel} 
\boldsymbol{\cdot} \nabla_{gy} F \\
&- e \frac{c \hat{b}}{e B^*_{\parallel}} \times \nabla \left < \phi_1 \right > 
\boldsymbol{\cdot} \nabla_{gy} F \\
&+ \frac{\textbf{B}^{*}}{m B_{\parallel}^{*}} 
\boldsymbol{\cdot} \left( e \nabla \left < \phi_1 \right > \right) 
\frac{\partial F}{\partial {v}_{gy, \parallel}} = 0.
\end{split}
\label{gyrokinetic_eq}
\end{equation}

Considering that $\textbf{B}^* = \textbf{B} + \frac{mc}{e} \bar{v}_{gy, \parallel} \nabla \times \hat{b} - \frac{mc^2}{e^2} \mu_{gy} \nabla ^* R^*$, and unless otherwise explicit, considering also that

\begin{equation}
    \frac{\textbf{B}^*}{B^*_\parallel} =  \frac{\textbf{B}^*}{B + \frac{m}{e}c {v}_{gy, \parallel} (\hat{b} \boldsymbol{\cdot} \nabla \times \hat{b}) - \frac{mc^2}{e^2} \mu_{gy} (\hat{b} \boldsymbol{\cdot} \nabla^*R^*)} \approx \hat{b},
\end{equation}

we have

\begin{equation*}
\frac{\partial F}{\partial t}
+ v_{gy,\parallel}\,\partial_{z,gy}F
- e\,\frac{c\hat{b}}{eB^*_{\parallel}}\times \partial_z\left<\phi_{1}\right>\,\boldsymbol{\cdot}\,\partial_{z,gy}F
\end{equation*}

\begin{equation*}
+\left(\frac{e}{m}\partial_z\left<\phi_{1}\right>\right)\partial_{v_{gy,\parallel}}F \, 
\end{equation*}

which means we can write our main Vlasov equations as 
  
\begin{equation}
\begin{split}
\frac{\partial F}{\partial t} 
&+ {v}_{gy, \parallel} \partial_{z,gy} F 
- \frac{c}{B_{\parallel}} 
\left( \hat{b} \times \partial_z \left< \phi_{1} \right> \right) 
\boldsymbol{\cdot} \partial_{z,gy} F \\
&+ \left(\frac{e}{m} \partial_z \left< \phi_{1} \right>\right) 
\partial_{v_{gy, \parallel}} F = 0.
\end{split}
\label{drift_kinetic_vlasov}
\end{equation}

Now we perform a linearization of the Vlasov and field equations, and proceed with the study of the modified electrostatic dispersion solutions of the FIDEL code \citep{Felix}.

\subsection{Driftkinetic electrons}

We proceed with the linearization of the drift kinetic Vlasov equation, which is responsible for the electron dynamics. For pedagogic reasons, we start the linearization from the very first equation \ref{25}, we have then 

\begin{equation}
\begin{split}
\frac{\partial F_e}{\partial t} 
&+ {v}_{gy, \parallel} \boldsymbol{\cdot} \nabla_{gy} F_e 
- \frac{c}{B^*_{\parallel}} 
\left( {{{\hat{b}}}} \times \nabla\left<\phi_{1}\right> \right)
\boldsymbol{\cdot} \nabla_{gy} F_e \\
&+ \left(\frac{e}{m_e} {\hat{b}} \boldsymbol{\cdot} \nabla\left<\phi_{1}\right>\right)
\partial_{v_{gy, \parallel}} F_e = 0.
\end{split}
\label{drift_kinetic_eq}
\end{equation}
  
The linearization is performed by working independently with each one of the terms in equation \ref{drift_kinetic_vlasov}.  First, we start with the decomposition of the distribution function, such as,
 \begin{equation*}
\frac{\partial F}{\partial t} = \frac{\partial}{\partial t} \left ( F_0 + \delta F \right ) = -i\omega \delta F,
\end{equation*}

where we considered $F = F_0 + \delta F e^{i(\textbf{k} \boldsymbol{\cdot} \textbf{x} - wt)}$ and $F_0$ to be a spatially homogeneous Maxwellian. Using the parallel component of $\nabla$ on the canonical moment, we have

 \begin{equation*}
\frac{\textbf{B}^*}{mB_\parallel^*} m {v}_{gy, \parallel} \nabla_{\parallel gy} \delta F_e =  \frac{\textbf{B}^*}{B_\parallel^*} {v}_{gy, \parallel} i k_\parallel \delta F_e,
\end{equation*}

and, due to the electrostatic nature of the present linearization, the second term on the first line of \ref{25} becomes
 \begin{equation*}
\frac{e\textbf{B}^*}{mcB_\parallel^*} \left < A_{1 \parallel }\right > \boldsymbol{\cdot} \nabla_{\parallel gy} \delta F_e = \frac{\Omega_{ce}}{B_\parallel ^ *} \delta A_{1 \parallel} i k_\parallel \delta F_e = 0.
\end{equation*}

The first term of the second line of \ref{25} is related to the $\nabla B$  drift, and becomes 

\[
\frac{c \hat{b}}{e B_\parallel} \times \mu_{gy} \nabla_{gy} B(X_{gy}) 
\boldsymbol{\cdot} \nabla_{gy} F_e
\]
\begin{equation}
\begin{split}
= \frac{c \hat{b}}{e B_\parallel} \times \mu_{gy} \nabla_{gy} B(X_{gy}) 
\boldsymbol{\cdot} \nabla_{\perp gy} \delta F_e = 0.
\end{split}
\label{gyrokinetic_equation}
\end{equation}

The $E \times B$ term becomes 

\begin{equation}
\frac{c \hat{b}}{e B_\parallel} \times e \nabla \left < \phi_1 \right > \boldsymbol{\cdot} \nabla_{gy}F_e = \frac{c}{B_\parallel} \hat{b} \times \nabla_\perp \phi_1 \boldsymbol{\cdot} \nabla_\perp \delta F_e =  0,
\end{equation}

and the magnetic flutter develops into

\begin{equation}
\frac{c \hat{b}}{e B_\parallel} \times \frac{e}{c} {v}_{gy \parallel} \nabla \left < A_{1 \parallel} \right > \boldsymbol{\cdot} \nabla_{gy}F_e = \frac{\hat{b}}{B_\parallel} \times  {v}_{gy \parallel} \nabla_\perp A_{1 \parallel} \boldsymbol{\cdot} \nabla_\perp \delta F_e = 0.
\end{equation}

The last line on equation \ref{25} is related to the parallel acceleration. For the first term on the same line we have that

\begin{equation*}
\frac{1}{m} \mu_{gy} \frac{\partial}{\partial z} B(X_{gy}) \partial_{v_{gy,\parallel}}F_e = 0.
\end{equation*}

The second term in the last line of \ref{25} becomes 
\begin{equation}
\begin{split}
\frac{\textbf{B}}{m B^*_\parallel} \boldsymbol{\cdot} 
e \nabla \left< \phi_1 \right> \partial_{v_{gy,\parallel}}F_e 
&= \frac{e \textbf{B}}{m B^*_{\parallel}} \boldsymbol{\cdot} 
i k \delta \phi \partial_{v_{gy,\parallel}}F_{0,e} \\
&= \frac{e}{m} i k_\parallel \delta \phi 
\partial_{v_{gy,\parallel}}F_{0,e}.
\end{split}
\label{gyrokinetic_response}
\end{equation}

and in the electrostatic limit, the third term becomes zero. It is also important to note that for the study of solar wind turbulence, a slab approximation is sufficient to model the geometry of the magnetic field. Consequently, the final linearized equation is

\begin{equation}
 -i\omega \delta F_e +  {v}_{gy, \parallel} i k_\parallel \delta F_e + \frac{e}{m} i{k_\parallel}\delta\phi\partial_{v_{gy,\parallel}}F_{0,e} = 0.\label{electron_vlasov_linear}
\end{equation}
The distribution function is computed as solutions of the system, and its integral gives us the electron gyrocenter density perturbation, $\delta n_e$, which will in turn be used in the linearized Poisson equation. For further details, we again refer the reader to  \citep{HYDROS, Nathan, Felix}.

\subsection{Vlasov ions}

We start with the linearization of the ion Vlasov equation. From equation \ref{final_fully} we have that

\begin{equation}
\frac{\partial F}{\partial t}+ \textbf{v}  \boldsymbol{\cdot} \nabla F+\frac{e}{m_{}}\left(\nabla \phi +\textbf{v} \times \textbf{B} \right) \boldsymbol{\cdot} \nabla_{v}F=0\label{ion_vlasov_1st_apndx}.
\end{equation}

We linearise equation \ref{ion_vlasov_1st_apndx} term-by-term, and then proceed to perform a plane wave assumption, such as 

\begin{equation}
 F = F_0 + \delta F e^{i \left( \textbf{k} \boldsymbol{\cdot} \textbf{x} -\omega t\right)}.
\end{equation}
The first term becomes then 

\begin{equation}
\frac{\partial F}{\partial t}=\frac{\partial}{\partial t}\left(F_{0}+\delta F\right)=-i\omega\delta F.
\end{equation}
The second and third terms become then

\begin{equation}
{\textbf{v}}\boldsymbol{\cdot}\nabla F=i{\textbf{v}\boldsymbol{\cdot}\boldsymbol{k}}\delta F,
\end{equation}
\begin{equation}
-\frac{e}{m}\nabla\phi\boldsymbol{\cdot}\nabla_{v}F=\frac{e}{m}i\boldsymbol{k}\delta\phi\boldsymbol{\cdot}\nabla_{v}F_{0}.
\end{equation}

Finally, the last term becomes

\begin{equation}
\textbf{v} \times \textbf{B} \boldsymbol{\cdot} \nabla_{v}\delta F = 0 .
\end{equation}

Considering all terms together, and considering a standard Maxwellian distribution for $F_0$, our linearized ion Vlasov equation takes the form

\begin{equation}
-i\omega\delta F + i{\textbf{v}\boldsymbol{\cdot}\boldsymbol{k}}\delta F -\frac{e}{m}i\boldsymbol{k}\delta\phi\boldsymbol{\cdot}\nabla_{v}F_{0}=0.\label{ion_vlasov_linear}
\end{equation}
The solutions for this system are integrated and we can then construct the linear ion density perturbation $\delta n_i$. The solution for this equation is rather complex and its derivation rather cumbersome, for details, we refer the reader to \cite{HYDROS, Felix}, and more specifically to \cite{Nathan}.

\subsection{Poisson equation}
To perform the linearization of the Poisson equation, we work with the following  considerations:

\begin{equation}
\frac{1}{4\pi}\nabla^{2}\phi_{1}\left(\textbf{x}\right)+\frac{\rho_{th}^{2}}{\lambda_{D}^{2}}\nabla_{\perp}^{2}\psi_{1}\left(\textbf{x}\right)=\sum_{i}q_{i} n _{i}\left(\textbf{x}\right)-e n _{e}\left(\textbf{x}\right),
\label{apndx_poisson}
\end{equation}

\begin{equation*}
\psi_{1}\left(\textbf{X}_{gy}+\boldsymbol{\rho}\right) \sim \phi_{1},
\end{equation*}
\begin{equation*}
\phi_{1}\left(x\right)=\tilde{\phi_{1}}\exp\left[i\left(\textbf{k}\boldsymbol{\cdot}\textbf{x}-\omega t\right)\right].
\end{equation*}

The left-hand side of equation \ref{apndx_poisson} becomes

\begin{equation}
\frac{1}{4\pi}\nabla^{2}\phi_{1}\left(x\right)+\frac{\rho_{th}^{2}}{\lambda_{D}^{2}}\nabla_{\perp}^{2}\psi_{1}\left(x\right)=
\end{equation}
\[
\frac{1}{4\pi}\left(k^{2}\tilde{\phi_{1}}\exp\left[i\left(\textbf{k}\boldsymbol{\cdot}\textbf{x}-\omega t\right)\right]\right)+ \frac{\rho_{th}^{2}}{\lambda_{D}^{2}}\left(-k_{\perp}^{2}\tilde{\phi_{1}}\exp\left[i\left(\textbf{k}\boldsymbol{\cdot}\textbf{x}-\omega t\right)\right] \right)
\]

\[
=\left(\frac{1}{4\pi}k^2+\frac{\rho_{th}^{2}}{\lambda_{D}^{2}}k_{\perp}^{2}\right)\Tilde{\phi}_{1}.
\label{eq102}
\]

Due to background quasi neutrality, the right side of equation (\ref{apndx_poisson}) can be expressed as

\begin{equation*}
\sum_{i}q_{i} n _{i}\left(x\right)+e n _{e}\left(x\right) = q_i \delta n_i - e \delta n_e,
\end{equation*}
and our  linearized Poisson equations becomes

\begin{equation}
\left( k^2 +\frac{1}{2}\frac{\omega_{pe}^2}{\Omega_{ce}^2}k_\perp^2\right)\phi_1=4\pi\left(q_i \delta n_i-e \delta n_e\right),
\label{poisson_linear}
\end{equation}
where $\delta n_i$ is the perturbed ion density and $ \delta n_{e}$ is the perturbed electron density computed in the guiding center position, also known as the gyrocenter density distribution. 

The equations \ref{ion_vlasov_linear}, \ref{electron_vlasov_linear}, and \ref{poisson_linear} constitute a set of linear equations that must be solved simultaneously. To find the solutions for the system, we first solve the perturbed ion and electron equations. By integrating these over velocity space, we can then solve the field equation given by \ref{poisson_linear}. This yields the wave dispersion relation, which can be expressed as

\begin{equation}
D(k,\omega) \delta \phi = 0,\label{D}
\end{equation}

where $D(k,\omega)$ is in general the determinant of the matrix of the system. The solutions of equation \ref{D} provide us with the various waves we aim to study.

\bibliography{your-bib-file}

@Article{Verscharen,
  author = 	 {Verscharen, D. and Klein, K.G. and Maruca, B.A.},
  title = 	 {The multi-scale nature of the solar wind},
  journal =  {Living Rev Sol Phys 16, 5},
  year = 	 {2019},
}

@Book{Carbone,
  author = 	 {Bruno, R. and  Carbone, V.},
  title = {},
  booktitle = 	 {Turbulence in the Solar Wind},
  publisher = {Springer},
  year = 	 {2016},
  editor = 	 {},
  volume = 	 {928},
  series = 	 {},
}

@Article{Was,
  author = 	 {Robert H. Wasserman},
  title = 	 {Tensor and Manifolds, with applications to Physics.},
  journal = 	 {Oxford},
  year = 	 {2004},
  volume = 	 {2nd Edition},
  pages = 	 {},
}

@article{heterogeneous_manifolds,
doi = {10.1088/2058-6272/ac18ba},
url = {https://dx.doi.org/10.1088/2058-6272/ac18ba},
year = {2021},
month = {aug},
publisher = {IOP Publishing},
volume = {23},
number = {10},
pages = {105103},
author = {Fan, Peifeng and Qin, Hong and Xiao, Jianyuan},
title = {Discovering exact, gauge-invariant, local energy–momentum conservation laws for the electromagnetic gyrokinetic system by high-order field theory on heterogeneous manifolds},
journal = {Plasma Science and Technology}
}

@Article{Told,
  author = 	 {Daniel Told and Patrick Astfalk and  Fabian Mueller and  Tessa Cookmayer and Frank Jenko},
  title = 	 {Comparative study of gyrokinetic, hybrid kinetic and fully kinetic wave physics for space plasmas.},
  journal = 	 {New Journal of Physics},
  year = 	 {2016},
  volume = 	 {16(6), 065011},
  pages = 	 {},
}

@Article{Littlejohn,
  author = 	 {Robert G. Littlejohn},
  title = 	 {Hamiltonian perturbation theory in noncanonical coordinates.},
  journal = 	 {Journal of Mathematical Physics},
  year = 	 {1982},
  volume = 	 {23(5)},
  pages = 	 {},
}

@article{littlejohn_1983, 
title={Variational principles of guiding centre motion}, 
volume={29}, 
DOI={10.1017/S002237780000060X}, 
number={1}, 
journal={Journal of Plasma Physics},
publisher={Cambridge University Press}, 
author={Littlejohn, Robert G.},
year={1983}, 
pages={111–125}
}

@Article{Sugama,
  author = 	 {H. Sugama},
  title = 	 {Gyrokinetic Field Theory.},
  journal = 	 {Physics of Plasmas},
  year = 	 {2000},
  volume = 	 {7(466)},
  pages = 	 {},
  }

@book{Lanczos,
    title = {The Variational Principle of Mechanics},
    author = {Cornelius Lanczos},
    isbn = {},
    series = {Landau lectures on physics},
    year = {1970},
    publisher = {Dover}
}

@Article{PSP_1,
  author = 	 {Bale, S.D. and Badman, S.T. and Bonnell, J.W.},
  title = 	 {Highly structured slow solar wind emerging from an equatorial coronal hole.},
  journal = 	 {Nature},
  year = 	 {2019},
  volume = 	 {576},
  pages = 	 { 237–242},
}

@article{PSP_2,
doi = {10.3847/1538-4365/ab4ff1},
url = {https://dx.doi.org/10.3847/1538-4365/ab4ff1},
year = {2020},
month = {feb},
publisher = {The American Astronomical Society},
volume = {246},
number = {2},
pages = {26},
author = {Zhao, L.-L. and Zank, G. P. and Adhikari, L. and Hu, Q. and Kasper, J. C. and Bale, S. D. and Korreck, K. E. and Case, A. W. and Stevens, M. and Bonnell, J. W. and Dudok de Wit, T. and Goetz, K. and Harvey, P. R. and MacDowall, R. J. and Malaspina, D. M. and Pulupa, M. and Larson, D. E. and Livi, R. and Whittlesey, P. and Klein, K. G.},
title = {Identification of Magnetic Flux Ropes from Parker Solar Probe Observations during the First Encounter},
journal = {The Astrophysical Journal Supplement Series}
}

@article{PSP_3,
doi = {10.3847/1538-4365/ab6c65},
url = {https://dx.doi.org/10.3847/1538-4365/ab6c65},
year = {2020},
month = {feb},
publisher = {The American Astronomical Society},
volume = {246},
number = {2},
pages = {66},
author = {Bowen, Trevor A. and Mallet, Alfred and Huang, Jia and Klein, Kristopher G. and Malaspina, David M. and Stevens, Michael and Bale, Stuart D. and Bonnell, J. W. and Case, Anthony W. and Chandran, Benjamin D. G. and Chaston, C. C. and Chen, Christopher H. K. and Dudok de Wit, Thierry and Goetz, Keith and Harvey, Peter R. and Howes, Gregory G. and Kasper, J. C. and Korreck, Kelly E. and Larson, Davin and Livi, Roberto and MacDowall, Robert J. and McManus, Michael D. and Pulupa, Marc and Verniero, J. L. and Whittlesey, Phyllis and (The PSP/FIELDS and PSP/SWEAP Teams)},
title = {Ion-scale Electromagnetic Waves in the Inner Heliosphere},
journal = {The Astrophysical Journal Supplement Series}
}

@article{PSP_4,
doi = {10.3847/1538-4365/ab64e6},
url = {https://dx.doi.org/10.3847/1538-4365/ab64e6},
year = {2020},
month = {feb},
publisher = {The American Astronomical Society},
volume = {246},
number = {2},
pages = {58},
author = {Parashar, T. N. and Goldstein, M. L. and Maruca, B. A. and Matthaeus, W. H. and Ruffolo, D. and Bandyopadhyay, R. and Chhiber, R. and Chasapis, A. and Qudsi, R. and Vech, D. and Roberts, D. A. and Bale, S. D. and Bonnell, J. W. and de Wit, T. Dudok and Goetz, K. and Harvey, P. R. and MacDowall, R. J. and Malaspina, D. and Pulupa, M. and Kasper, J. C. and Korreck, K. E. and Case, A. W. and Stevens, M. and Whittlesey, P. and Larson, D. and Livi, R. and Velli, M. and Raouafi, N.},
title = {Measures of Scale-dependent Alfvénicity in the First PSP Solar Encounter},
journal = {The Astrophysical Journal Supplement Series}
}

@article{PSP_5,
doi = {10.3847/1538-4365/ab6220},
url = {https://dx.doi.org/10.3847/1538-4365/ab6220},
year = {2020},
month = {feb},
publisher = {The American Astronomical Society},
volume = {246},
number = {2},
pages = {61},
author = {Bandyopadhyay, Riddhi and Matthaeus, W. H. and Parashar, T. N. and Chhiber, R. and Ruffolo, D. and Goldstein, M. L. and Maruca, B. A. and Chasapis, A. and Qudsi, R. and McComas, D. J. and Christian, E. R. and Szalay, J. R. and Joyce, C. J. and Giacalone, J. and Schwadron, N. A. and Mitchell, D. G. and Hill, M. E. and Wiedenbeck, M. E. and McNutt, R. L. and Desai, M. I. and Bale, Stuart D. and Bonnell, J. W. and de Wit, Thierry Dudok and Goetz, Keith and Harvey, Peter R. and MacDowall, Robert J. and Malaspina, David M. and Pulupa, Marc and Velli, M. and Kasper, J. C. and Korreck, K. E. and Stevens, M. and Case, A. W. and Raouafi, N.},
title = {Observations of Energetic-particle Population Enhancements along Intermittent Structures near the Sun from the Parker Solar Probe},
journal = {The Astrophysical Journal Supplement Series}
}

@article{PSP_6,
doi = {10.3847/1538-4365/ab74e0},
url = {https://dx.doi.org/10.3847/1538-4365/ab74e0},
year = {2020},
month = {mar},
publisher = {The American Astronomical Society},
volume = {246},
number = {2},
pages = {70},
author = {Huang, Jia and Kasper, J. C. and Vech, D. and Klein, K. G. and Stevens, M. and Martinović, Mihailo M. and Alterman, B. L. and Ďurovcová, Tereza and Paulson, Kristoff and Maruca, Bennett A. and Qudsi, Ramiz A. and Case, A. W. and Korreck, K. E. and Jian, Lan K. and Velli, Marco and Lavraud, B. and Hegedus, A. and Bert, C. M. and Holmes, J. and Bale, Stuart D. and Larson, Davin E. and Livi, Roberto and Whittlesey, P. and Pulupa, Marc and MacDowall, Robert J. and Malaspina, David M. and Bonnell, John W. and Harvey, Peter and Goetz, Keith and de Wit, Thierry Dudok},
title = {Proton Temperature Anisotropy Variations in Inner Heliosphere Estimated with the First Parker Solar Probe Observations},
journal = {The Astrophysical Journal Supplement Series}
}

@article{PSP_7,
doi = {10.3847/1538-4365/ab5853},
url = {https://dx.doi.org/10.3847/1538-4365/ab5853},
year = {2020},
month = {feb},
publisher = {The American Astronomical Society},
volume = {246},
number = {2},
pages = {39},
author = {Dudok de Wit, Thierry and Krasnoselskikh, Vladimir V. and Bale, Stuart D. and Bonnell, John W. and Bowen, Trevor A. and Chen, Christopher H. K. and Froment, Clara and Goetz, Keith and Harvey, Peter R. and Jagarlamudi, Vamsee Krishna and Larosa, Andrea and MacDowall, Robert J. and Malaspina, David M. and Matthaeus, William H. and Pulupa, Marc and Velli, Marco and Whittlesey, Phyllis L.},
title = {Switchbacks in the Near-Sun Magnetic Field: Long Memory and Impact on the Turbulence Cascade},
journal = {The Astrophysical Journal Supplement Series}
}

@article{PSP_8,
doi = {10.3847/1538-4365/ab527f},
url = {https://dx.doi.org/10.3847/1538-4365/ab527f},
year = {2020},
month = {feb},
publisher = {The American Astronomical Society},
volume = {246},
number = {2},
pages = {30},
author = {Martinović, Mihailo M. and Klein, Kristopher G. and Kasper, Justin C. and Case, Anthony W. and Korreck, Kelly E. and Larson, Davin and Livi, Roberto and Stevens, Michael and Whittlesey, Phyllis and Chandran, Benjamin D. G. and Alterman, Ben L. and Huang, Jia and Chen, Christopher H. K. and Bale, Stuart D. and Pulupa, Marc and Malaspina, David M. and Bonnell, John W. and Harvey, Peter R. and Goetz, Keith and Dudok de Wit, Thierry and MacDowall, Robert J.},
title = {The Enhancement of Proton Stochastic Heating in the Near-Sun Solar Wind},
journal = {The Astrophysical Journal Supplement Series}
}

@ARTICLE{PSP_9,
       author = {{Chen}, C.~H.~K. and et al},
        title = "{The Evolution and Role of Solar Wind Turbulence in the Inner Heliosphere}",
      journal = {arXiv e-prints},
         year = "2019",
        month = "Dec",
          eid = {arXiv:1912.02348},
        pages = {arXiv:1912.02348},
archivePrefix = {arXiv},
       eprint = {1912.02348},
 primaryClass = {astro-ph.SR},
       adsurl = {https://ui.adsabs.harvard.edu/abs/2019arXiv191202348C}
}

@article{Alexandrova2009,
  title = {Universality of Solar-Wind Turbulent Spectrum from MHD to Electron Scales},
  author = {Alexandrova, O. and Saur, J. and Lacombe, C. and Mangeney, A. and Mitchell, J. and Schwartz, S. J. and Robert, P.},
  journal = {Phys. Rev. Lett.},
  volume = {103},
  issue = {16},
  pages = {165003},
  numpages = {4},
  year = {2009},
  month = {Oct},
  publisher = {American Physical Society},
  doi = {10.1103/PhysRevLett.103.165003},
  url = {https://link.aps.org/doi/10.1103/PhysRevLett.103.165003}
}

@book{book_BasicAstro,
    title = {Basic of Plasma Astrophysics},
    author = {Claudio Chiuderi and Marco Velli},
    isbn = {978-88-470-5281-9},
    series = {},
    year = {2015},
    publisher = {Springer}
}

@book{book_SpacePhysics,
    title = {Space Physics, An Introduction to Plasmas and Particles in the Heliosphere and Magnetosphere.},
    author = {M.-B. Kallenrode},
    isbn = {},
    series = {Advanced Text in Physics},
    year = {2001},
    number = {},
    publisher = {Springer}
}

@book{book_PlasmaAstrophysics,
    title = {Plasma Astrophysics},
    author = {Kirk, J. G. and Melrose, D. B. and Priest, E. R.},
    isbn = {},
    series = {Lecture Notes 1994 of the Swiss Society for Astrophysics and Astronomy},
    year = {1994},
    number = {},
    publisher = {Springer-Verlag}
}

@book{book_TurbulenceWind,
    title = {Turbulence in the Solar Wind},
    author = {Bruno, R. and Carbone, V.},
    isbn = {},
    series = {Lecture Notes in Physics 928},
    year = {2016},
    number = {},
    publisher = {Springer}
}

@article{TurbulenceDissipation,
	doi = {10.3847/1538-4357/ab2a79},
	url = {},
	year = 2019,
	month = {aug},
	publisher = {American Astronomical Society},
	volume = {880},
	number = {2},
	pages = {121},
	author = {He, Jiansen and Duan, Die and Wang, Tieyan and Zhu, Xingyu and Li, Wenya and Verscharen, Daniel and Wang, Xin and Tu, Chuanyi and Khotyaintsev, Yuri and Le, Guan and Burch, Jim},
	title = {Direct Measurement of the Dissipation Rate Spectrum around Ion Kinetic Scales in Space Plasma Turbulence},
	journal = {The Astrophysical Journal}
}

@article{Chen2019,
	year = 2019,
	publisher = {Nature},
	volume = {740},
	number = {10},
	author = {Chen, C.H.K. and Klein, K.G. and Howes, G.G.},
	title = {Evidence for electron Landau damping in space plasma turbulence},
	journal = {Nature Communications},
}

@article{Howes2006,
archivePrefix = {arXiv},
arxivId = {astro-ph/0511812},
author = {Howes, G. G. and Cowley, S. C. and Dorland, W. and Hammett, G. W. and Quataert, E. and Schekochihin, A. A.},
doi = {10.1086/506172},
eprint = {0511812},
issn = {0004-637X},
journal = {The Astrophysical Journal},
number = {1},
pages = {590--614},
primaryClass = {astro-ph},
title = {{Astrophysical Gyrokinetics: Basic Equations and Linear Theory}},
volume = {651},
year = {2006}
}

@article{Tronko2018,
archivePrefix = {arXiv},
arxivId = {1709.05222},
author = {Tronko, Natalia and Chandre, Cristel},
doi = {10.1017/S0022377818000430},
eprint = {1709.05222},
issn = {14697807},
journal = {Journal of Plasma Physics},
number = {3},
title = {{Second-order nonlinear gyrokinetic theory: From the particle to the gyrocentre}},
volume = {84},
year = {2018}
}

@article{metriplectic,
title = {Energetically consistent model reduction for metriplectic systems},
journal = {Computer Methods in Applied Mechanics and Engineering},
volume = {404},
pages = {115709},
year = {2023},
issn = {0045-7825},
doi = {https://doi.org/10.1016/j.cma.2022.115709},
url = {https://www.sciencedirect.com/science/article/pii/S0045782522006648},
author = {Anthony Gruber and Max Gunzburger and Lili Ju and Zhu Wang}
}

@book{MHD_1,
    title = {An Introduction to Plasma Astrophysics and Magnetohydrodynamics},
    author = {M. Goossens},
    isbn = {1402014295},
    series = {},
    year = {2003},
    publisher = {Springer Netherlands}
}

@book{MHD_2,
    title = {Magnetohydrodynamics of Laboratory and Astrophysical Plasmas},
    author = {Hans Goedbloed, Rony Keppens, Stefaan Poedts},
    isbn = {9781107123922},
    series = {},
    year = {2019},
    publisher = {Cambridge University Proess}
}

@article{KAW_1,
	doi = {10.3847/2041-8213/ab2fe6},
	url = {https://doi.org/10.3847\%2F2041-8213\%2Fab2fe6},
	year = {2019},
	month = {jul},
	publisher = {American Astronomical Society},
	volume = {880},
	number = {1},
	pages = {L10},
	author = {Vincent David and S{\'{e}}bastien Galtier},
	title = {$k_\perp^{-8/3}$ Spectrum in Kinetic Alfv{\'{e}}n Wave Turbulence: Implications for the Solar Wind},
	journal = {The Astrophysical Journal}
}

@book{Callen,
    title = {Thermodynamics and an Introduction to Thermostatistics},
    author = {Herbert B. Callen},
    isbn = {},
    series = {},
    year = {1985},
    publisher = {John Wiley and Sons}
}

@article{Viscous,
	doi = {},
	url = {},
	year = {2008},
	month = {},
	publisher = {Springer},
	volume = {},
	number = {},
	pages = {},
	author = { G.L. Morini},
	title = {Viscous Dissipation},
	journal = {Li D. (eds) Encyclopedia of Microfluidics and Nanofluidics},
}

@article{Relaxation,
  title = {Collisional Relaxation of Fine Velocity Structures in Plasmas},
  author = {Pezzi, Oreste and Valentini, Francesco and Veltri, Pierluigi},
  journal = {Phys. Rev. Lett.},
  volume = {116},
  issue = {14},
  pages = {145001},
  numpages = {5},
  year = {2016},
  month = {Apr},
  publisher = {American Physical Society},
  doi = {10.1103/PhysRevLett.116.145001},
  url = {https://link.aps.org/doi/10.1103/PhysRevLett.116.145001}
}

@article{Relaxation2,
	doi = {10.3847/1538-4357/835/2/133},
	url = {https://doi.org/10.3847%2F1538-4357%2F835%2F2%2F133},
	year = 2017,
	month = {jan},
	publisher = {American Astronomical Society},
	volume = {835},
	number = {2},
	pages = {133},
	author = {Schreiner, A. and Saur, J.},
	title = {A Model for Dissipation of Solar Wind Magnetic Turbulence by Kinetic Alfv{\'{e}}n Waves at Electron Scales: Comparison with Observations},
	journal = {The Astrophysical Journal}
}

@book{pitaevskii2012physical,
  title={Physical Kinetics},
  author={Pitaevskii, L.P. and Lifshitz, E.M.},
  number={Bd. 10},
  series = {0},
  isbn={9780080570495},
  lccn={80042162},
  url={https://books.google.de/books?id=DTHxPDfV0fQC},
  year={2012},
  publisher={Elsevier Science}
}

@article{CyclotronDamping,
author = {Olson,Craig L. },
title = {Spatial Electron Cyclotron Damping},
journal = {The Physics of Fluids},
volume = {15},
number = {1},
pages = {160-165},
year = {1972},
doi = {10.1063/1.1693733},

URL = { 
        https://aip.scitation.org/doi/abs/10.1063/1.1693733
    
},
eprint = { 
        https://aip.scitation.org/doi/pdf/10.1063/1.1693733
    
}

}

@article{Stochastic,
	doi = {10.3847/1538-4357/ab23f4},
	url = {https://doi.org/10.3847%2F1538-4357%2Fab23f4},
	year = 2019,
	month = {jul},
	publisher = {American Astronomical Society},
	volume = {879},
	number = {1},
	pages = {43},
	author = {Mihailo M. Martinovi{\'{c}} and Kristopher G. Klein and Sofiane Bourouaine},
	title = {Radial Evolution of Stochastic Heating in Low-beta Solar Wind},
	journal = {The Astrophysical Journal},
}

@article{Pumping,
title = "Theory of collisional heating of plasma by magnetic pumping",
journal = "Physica",
volume = "32",
number = "8",
pages = "1410 - 1428",
year = "1966",
issn = "0031-8914",
doi = "https://doi.org/10.1016/0031-8914(66)90133-9",
url = "http://www.sciencedirect.com/science/article/pii/0031891466901339",
author = "R. Gerwin"
}

@article{Contact,
title = "Contact geometry: The geometrical method of Gibbs’s thermodynamics",
journal = "Proc. Gibbs Symp",
pages = "163-179",
year = "1990",
author = "V.I. Arnold",
}

@article{Low,
author = {Low, F. E.  and Chandrasekhar, S. },
title = {A Lagrangian formulation of the Boltzmann-Vlasov equation for plasmas},
journal = {Proceedings of the Royal Society of London. Series A. Mathematical and Physical Sciences},
volume = {248},
number = {1253},
pages = {282-287},
year = {1958},
doi = {10.1098/rspa.1958.0244}
}

@article{Marsch,
author = {Marsch, E. and C. K. Goertz and K. Richter},
title = {Wave heating and acceleration of solar wind ions by cyclotron resonance},
journal = {J. Geophys. Res.},
volume = {87},
number = {5030},
year = {1982},
}

@article{Kohl,
doi = {10.1086/311434},
url = {https://dx.doi.org/10.1086/311434},
year = {1998},
month = {jun},
publisher = {},
volume = {501},
number = {1},
pages = {L127},
author = {Kohl, J. L. and Noci, G. and Antonucci, E. and Tondello, G. and Huber, M. C. E. and Cranmer, S. R. and Strachan, L. and Panasyuk, A. V and Gardner, L. D. and Romoli, M. and Fineschi, S. and Dobrzycka, D. and Raymond, J. C. and Nicolosi, P. and Siegmund, O. H. W. and Spadaro, D. and Benna, C. and Ciaravella, A. and Giordano, S. and Habbal, S. R. and Karovska, M. and Li, X. and Martin, R. and Michels, J. G. and Modigliani, A. and Naletto, G. and O'Neal, R. H. and Pernechele, C. and Poletto, G. and Smith, P. L. and Suleiman, R. M.},
title = {UVCS/SOHO Empirical Determinations of Anisotropic Velocity Distributions in the Solar Corona},
journal = {The Astrophysical Journal}
}

@masterthesis{Felix,
  author = {F. Gaisbauer},
  title = {A linear dispersion relation for a hybrid plasma model of drift-kinetic electrons and fully kinetic ions. Master thesis},
  school = {LMU},
  year = {2018},
  address = {}
}

@article{DSHARK,
author = {Astfalk P, Görler T and Jenko F},
title = {DSHARK: a dispersion relation solver for obliquely propagating waves in bi-kappa-distributed plasmas},
journal = {J. Geophys. Res},
volume = {120},
number = {7107–20},
year = {2015},
}

@article{HYDROS,
doi = {10.1088/1367-2630/18/7/075001},
url = {https://dx.doi.org/10.1088/1367-2630/18/7/075001},
year = {2016},
month = {jul},
publisher = {IOP Publishing},
volume = {18},
number = {7},
pages = {075001},
author = {Told, D and Cookmeyer, J and Astfalk, P and Jenko, F},
title = {A linear dispersion relation for the hybrid kinetic-ion/fluid-electron model of plasma physics},
journal = {New Journal of Physics}
}

@article{PhysRevResearch.2.043253,
  title = {Possible coexistence of kinetic Alfven and ion Bernstein modes in sub-ion scale compressive turbulence in the solar wind},
  author = {Roberts, Owen Wyn and Verscharen, Daniel and Narita, Yasuhito and Nakamura, Rumi and V\"or\"os, Zolt\'an and Plaschke, Ferdinand},
  journal = {Phys. Rev. Research},
  volume = {2},
  issue = {4},
  pages = {043253},
  numpages = {13},
  year = {2020},
  month = {Nov},
  publisher = {American Physical Society},
  doi = {10.1103/PhysRevResearch.2.043253},
  url = {https://link.aps.org/doi/10.1103/PhysRevResearch.2.043253}
}

@book{Plasma_Astro,
  title={Plasma Astrophysics, Part I},
  author={Boris V. Somov},
  number={},
  isbn={},
  lccn={},
  url={},
  year={2006},
  publisher={Springer-Verlag New York}
}

@article{IAW_1,
title = {Ion acoustic waves in the plasma with the power-law q-distribution in nonextensive statistics},
journal = {Physica A: Statistical Mechanics and its Applications},
volume = {387},
number = {19},
pages = {4821-4827},
year = {2008},
issn = {0378-4371},
doi = {https://doi.org/10.1016/j.physa.2008.04.016},
url = {https://www.sciencedirect.com/science/article/pii/S0378437108003907},
author = {Liu Liyan and Du Jiulin}
}

@article{IAW_3,
author = {D. A. G Urnnet and L. A. Frank },
title = {Ion Acoustic Waves in the Solar Wind},
journal = {Journal of Geophysical Research},
volume = {83},
number = {A1},
pages = {},
year = {1978},
}

@article{IAW_4,
author = {Paul J. Kellogg },
title = {Heating of the Solar Wind by Ion Acoustic Waves},
journal ={The Astrophysical Journal},
volume = {891 51},
number = {A1},
pages = {},
year = {2020},
doi = {10.3847/1538-4357/ab7003}
}

@article{IBW_1,
author = {Ono,Masayuki },
title = {Ion Bernstein wave heating research},
journal = {Physics of Fluids B: Plasma Physics},
volume = {5},
number = {2},
pages = {241-280},
year = {1993},
doi = {10.1063/1.860569},
URL = { https://doi.org/10.1063/1.860569},
eprint = { https://doi.org/10.1063/1.860569}
}

@article{IBW_2,
author = {Podesta, John J.},
title = {The need to consider ion Bernstein waves as a dissipation channel of solar wind turbulence},
journal = {Journal of Geophysical Research: Space Physics},
volume = {117},
number = {A7},
pages = {},
year = {2012}
}

@ARTICLE{high_freq_1,
       author = {Coleman, Paul J., Jr.},
        title = "{Turbulence, Viscosity, and Dissipation in the Solar-Wind Plasma}",
      journal = {Astrophysical Journal},
         year = {1968},
        month = {aug},
       volume = {153},
        pages = {371},
          doi = {10.1086/149674},
       adsurl = {https://ui.adsabs.harvard.edu/abs/1968ApJ...153..371C}
}

@incollection{high_freq_2,
title = {Coronal holes and the solar wind},
editor = {Petrus C.H. Martens and David P. Cauffman},
series = {COSPAR Colloquia Series},
publisher = {Pergamon},
volume = {13},
pages = {3-12},
year = {2002},
booktitle = {Multi-wavelength Observations of Coronal Structure and Dynamics},
issn = {0964-2749},
doi = {https://doi.org/10.1016/S0964-2749(02)80003-8},
url = {https://www.sciencedirect.com/science/article/pii/S0964274902800038},
author = {S.R. Cranmer}
}

@article{high_freq_3,
author = {Oughton, S. and Matthaeus, W. H. and Smith, C. W. and Breech, B. and Isenberg, P. A.},
title = {Transport of solar wind fluctuations: A two-component model},
journal = {Journal of Geophysical Research: Space Physics},
volume = {116},
number = {A8},
pages = {},
year = {2011}
}

@article{high_freq_4,
    author = {Karimabadi, H. and Roytershteyn, V. and Vu, H. X. and Omelchenko, Y. A. and Scudder, J. and Daughton, W. and Dimmock, A. and Nykyri, K. and Wan, M. and Sibeck, D. and Tatineni, M. and Majumdar, A. and Loring, B. and Geveci, B.},
    title = {The link between shocks, turbulence, and magnetic reconnection in collisionless plasmas},
    journal = {Physics of Plasmas},
    volume = {21},
    number = {6},
    pages = {062308},
    year = {2014},
    month = {06},
    issn = {1070-664X},
}

@article{high_freq_5,
author = {Hollweg, Joseph V.},
title = {A quasi-linear WKB kinetic theory for nonplanar waves in a nonhomogeneous warm plasma, 1. Transverse waves propagating along axisymmetric B 0},
journal = {Journal of Geophysical Research: Space Physics},
volume = {83},
number = {A2},
pages = {563-574},
doi = {https://doi.org/10.1029/JA083iA02p00563},
year = {1978}
}

@article{Goldstein2001,
author = {Goldstein, Melvyn L},
doi = {10.1023/A:1012264131485},
issn = {1572-946X},
journal = {Astrophysics and Space Science},
number = {1},
pages = {349--369},
title = {{Major Unsolved Problems in Space Plasma Physics}},
url = {https://doi.org/10.1023/A:1012264131485},
volume = {277},
year = {2001}
}

@article{9_outstanding_problems,
author = {Viall, N. M. and Borovsky, J. E.},
title = {Nine Outstanding Questions of Solar Wind Physics},
journal = {Journal of Geophysical Research: Space Physics},
volume = {125},
number = {7},
year = {2020},
}

@article{Chandran2010,
author = {Chandran, Benjamin D.G. and Li, Bo and Rogers, Barrett N. and Quataert, Eliot and Germaschewski, Kai},
doi = {10.1088/0004-637X/720/1/503},
issn = {15384357},
journal = {Astrophysical Journal},
number = {1},
pages = {503--515},
title = {{Perpendicular ion heating by low-frequency Alfv{\'{e}}n-wave turbulence in the solar wind}},
volume = {720},
year = {2010}
}

@incollection{Frank-Kamenetskii1972,
author = {Frank-Kamenetskii, D. A.},
booktitle = {Plasma: The Fourth State of Matter},
doi = {10.1007/978-1-4684-1896-5_27},
pages = {86--89},
publisher = {Springer US},
title = {{The Magnetization Current}},
url = {https://link.springer.com/chapter/10.1007/978-1-4684-1896-5_27},
year = {1972}
}

@PhdThesis{Nathan,
  author = {deOliveira-Lopes, F. N.},
  title = {Geometrical Formulation of Hybrid Kinetic and Gyrokinetic Hamiltonian Field Theory for Astrophysical and Laboratory Plasmas},
  school = {Ruhr-Universität Bochum},
  year = {2022},
  address = {}
}

@article{Lin_2005,
doi = {10.1088/0741-3335/47/4/006},
url = {https://dx.doi.org/10.1088/0741-3335/47/4/006},
year = {2005},
month = {mar},
publisher = {},
volume = {47},
number = {4},
pages = {657},
author = {Yu Lin and Xueyi Wang and Zhihong Lin and Liu Chen},
title = {A gyrokinetic electron and fully kinetic ion plasma simulation model},
journal = {Plasma Physics and Controlled Fusion}
}

@article{slmp,
doi = {10.3847/1538-4357/aa901f},
url = {https://dx.doi.org/10.3847/1538-4357/aa901f},
year = {2017},
month = {nov},
publisher = {The American Astronomical Society},
volume = {849},
number = {2},
pages = {76},
author = {Tanaka, Satoshi and Yoshikawa, Kohji and Minoshima, Takashi and Yoshida, Naoki},
title = {Multidimensional Vlasov–Poisson Simulations with High-order Monotonicity- and Positivity-preserving Schemes},
journal = {The Astrophysical Journal}
}

@article{semi_lagrange,
title = {The Semi-Lagrangian Method for the Numerical Resolution of the Vlasov Equation},
journal = {Journal of Computational Physics},
volume = {149},
number = {2},
pages = {201-220},
year = {1999},
issn = {0021-9991},
doi = {https://doi.org/10.1006/jcph.1998.6148},
url = {https://www.sciencedirect.com/science/article/pii/S0021999198961484},
author = {Eric Sonnendrücker and Jean Roche and Pierre Bertrand and Alain Ghizzo}
}

@book{domain_decomposition,
    author = {Quarteroni, Alfio and Valli, Alberto},
    title = {Domain Decomposition Methods for Partial Differential Equations},
    publisher = {Oxford University Press},
    year = {1999},
    month = {05},
    isbn = {9780198501787},
    doi = {10.1093/oso/9780198501787.001.0001},
    url = {https://doi.org/10.1093/oso/9780198501787.001.0001},
}

@article{cheng1976,
  title        = {The integration of the Vlasov equation in configuration space},
  author       = {Cheng, C.Z. and Knorr, G.},
  journal      = {Journal of Computational Physics},
  volume       = {22},
  number       = {3},
  pages        = {330--351},
  year         = {1976},
  doi          = {10.1016/0021-9991(76)90053-X}
}

@article{sonnendrucker1999,
  title        = {The semi-Lagrangian method for the numerical resolution of the Vlasov equation},
  author       = {Sonnendrücker, E. and Roche, J. and Bertrand, P. and Ghizzo, A.},
  journal      = {Journal of Computational Physics},
  volume       = {149},
  number       = {2},
  pages        = {201--220},
  year         = {1999},
  doi          = {10.1006/jcph.1998.6148}
}

@article{filbet2001,
  title        = {Conservative numerical schemes for the Vlasov equation},
  author       = {Filbet, F. and Sonnendrücker, E. and Bertrand, P.},
  journal      = {Journal of Computational Physics},
  volume       = {172},
  number       = {1},
  pages        = {166--187},
  year         = {2001},
  doi          = {10.1006/jcph.2001.6818}
}

@article{crouseilles2010,
  title        = {Conservative semi-Lagrangian schemes for Vlasov equations},
  author       = {Crouseilles, N. and Mehrenberger, M. and Sonnendrücker, E.},
  journal      = {Journal of Computational Physics},
  volume       = {229},
  number       = {6},
  pages        = {1927--1953},
  year         = {2010},
  doi          = {10.1016/j.jcp.2009.11.007}
}

@article{liu2023,
  title        = {An energy-conserving conservative semi-Lagrangian scheme for the Vlasov--Ampère system},
  author       = {Liu, H. and Cai, X. and Cao, Y. and Lapenta, G.},
  journal      = {Journal of Computational Physics},
  volume       = {492},
  pages        = {112412},
  year         = {2023},
  doi          = {10.1016/j.jcp.2023.112412}
}

@article{chang2021,
  title   = {Highly accurate monotonicity-preserving Semi-Lagrangian scheme for Vlasov-Poisson simulations},
  author  = {Chang, Y. and Michel, M.},
  journal = {Journal of Computational Physics},
  volume  = {446},
  pages   = {110632},
  year    = {2021},
  doi     = {10.1016/j.jcp.2021.110632}
}

@article{thatikonda2025,
  author = {Thatikonda, S. and De Oliveira-Lopes, F. N. and Mustonen, A. and Pommois, K. and Told, D. and Jenko, F.},
  title = {Verification of a hybrid kinetic--gyrokinetic model using the advanced semi-Lagrange code ssV},
   journal = {Computer Physics Communications},
  volume = {320},
  number = {},
  pages = {109980},
  year = {2025},
doi={10.1016/j.cpc.2025.109980}
}

\end{document}